\documentstyle[seceq,epsf]{ptptex}

\newlength{\minitwocolumn}
\newlength{\minithreecolumnsep}
\newlength{\minithreecolumn}
\title{
Projection Operator Approach to Langevin Equations in $\phi^4$ Theory
}

\author{
Tomoi {\sc Koide}$^{*}$,Masahiro {\sc Maruyama} and Fujio {\sc Takagi} }

\inst{
$^{*}$Yukawa Institute for Theoretical Physics, Kyoto University,
Kyoto 606-8502, Japan \\
Department of Physics, Tohoku University, Sendai 980-8578, Japan
}

\recdate{
October 10, 2001
}

\abst{
We apply the projection operator method (POM) to $\phi^4$ theory and 
derive both quantum and semiclassical equations of motion for 
the soft modes.
These equations have no time-convolution integral term, in sharp contrast with 
other well-known results obtained using the influence functional method (IFM) 
and the closed time path method (CTP).
However, except for the fluctuation force field terms, these equations are 
similar to the corresponding equations obtained using IFM 
with the linear harmonic approximation, which was introduced 
to remove the time-convolution integral.
The quantum equation of motion in POM can be regarded as 
a kind of quantum Langevin equation in which 
the fluctuation force field is given 
in terms of the operators of the hard modes.
These operators are then replaced with c-numbers using a certain procedure 
to obtain a semiclassical Langevin equation.
It is pointed out that there are significant differences between 
the fluctuation force fields introduced in this paper and those 
introduced in IFM.
The arbitrariness of the definition 
of the fluctuation force field in IFM is also discussed.}

\begin{document}

\maketitle

\section{Introduction}

In recent years,
there have been many studies 
related to time dependent phenomena in quantum systems:
the time development of order parameters in phase transitions,
\cite{ref:Morikawa,ref:Glei-Ramo,ref:Ber-Glei-Ramo,ref:Lom-Mazz,ref:Grei-Mul,ref:Rish,ref:Biro-Grei,ref:Boya-Vega-Hol,ref:Boya-Att-Vega,ref:Coo-Klu-Mott-Paz,ref:Rand}
the experimental observation of non-exponential decay in quantum
tunneling,
\cite{ref:nonexp}
the absence of the quantum Zeno effect in quantum field theory,\cite{ref:noqze}
the time evolution of the Bose-Einstein condensate,\cite{ref:Mies,ref:Barci}
renormalization in time evolution calculations,
\cite{ref:Stu,ref:Bogo-Shir,ref:Koide1,ref:Koide2,ref:Baacke-Hei-Pat,ref:Baacke-Boya-Vega,ref:Shve}
and so on.
In such studies, it is often important to evaluate the time evolution 
within quantum field theory.
However, the equation of motion in quantum field theory
is a nonlinear operator equation,
and hence it is difficult to solve in general.
There have been several attempts to solve it approximately.

One attempt to approximately solve the equaiton of motion involves 
a derivation of a semiclassical equation of motion
that has a fluctuation force field term,
in other words, colored noise.
This equation can be regarded as a kind of a Langevin equation,
and it may be possible to incorporate
the effects of both quantum and thermal fluctuations
into it through not only the coefficients in the equation
but also the fluctuation force field.
The Langevin equation is a well-developed tool 
in nonequilibrium statistical mechanics, and there exists 
many techniques for treating it, including mode coupling theory
to calculate dynamical critical exponents,\cite{cri-dyn}
scaling theory of nonequilibrium systems
near instability points.\cite{scailing}
Furthermore,
the effect of the fluctuation force field 
in the nonlinear Langevin equation is an important topic of study 
in various fields of physics, e.g. stochastic resonance\cite{ref:sto}
and noise-induced phase transition.\cite{ref:N-I-PT}
Such effects have not been studied in detail in cases that 
quantum field theory is needed to describe the relevant phenomena,
for example the time evolution of a disoriented chiral condensate, 
which may be a signature of quark-gluon plasmas.
The Langevin equation is the starting point in the analysis of such phenomena.

The derivation of Langevin equations in quantum field theory has 
been carried out mainly on the basis of the influence functional method (IFM)
and the closed time path method (CTP).
\cite{ref:Morikawa,ref:Glei-Ramo,ref:Ber-Glei-Ramo,ref:Lom-Mazz,ref:Grei-Mul,ref:Rish}
The resultant Langevin equation has two notable characteristics.
First, it has in general a time-convolution integral term,
which is often called a memory term.
Such a term makes the Langevin equation difficult to solve.
For this reason, it is usually removed by employing 
the quasi-instantaneous approximation\cite{ref:Morikawa,ref:Glei-Ramo}
or the linear harmonic approximation.\cite{ref:Grei-Mul}
The meaning of the quasi-instantaneous approximation is very clear.
\cite{ref:Rau-Mul}
However, it is known that with this approximation there can be no 
dissipation effect (as, for example, in $\phi^4$ theory), and therefore 
an additional approximation is employed.
Contrastingly, with the linear harmonic approximation
there is a dissipation effect
in $\phi^4$ theory without the use of an additional approximation.
However, the validity of this approximation is not clear.
The second characteristic of the Langevin eqaution obtained using IFM 
and CTP is that the fluctuation force field is introduced 
using an auxiliary field.
It is not clear why the auxiliary field introduced 
in this way can be interpreted as 
the fluctuation force field.\cite{ref:Sch,ref:Cal-Legg}
Furthermore, the definition of the fluctuation force field is not unique, 
as is discussed in this paper.

The projection operator method (POM)
is another method to derive the Langevin equation for quantum systems.
\cite{ref:Mori,ref:Toku-Mori,ref:Naka-Zwa,ref:H-S-S,ref:S-H,ref:U-S,ref:Koide-Maruyama,ref:S-H-flu}
A particular version of this method was proposed 
by Hashitsume, Shibata and Shing\=u,\cite{ref:H-S-S,ref:S-H} 
and was improved by Uchiyama and Shibata\cite{ref:U-S}
and Koide and Maruyama\cite{ref:Koide-Maruyama} independently.
Using the Mori projection as the projection operator,
one can derive the Mori equation,
which is a well-known linear Langevin equation applicable to 
classical and quantum systems.
\cite{ref:Mori,ref:Toku-Mori}
POM is the only method that can be used to systematically 
derive an equation of motion 
without the time-convolution integral, as far as we know.
The fluctuation force field at the quantum level is explicitly given 
as a term representing the time variation 
in the space that is projected out, and 
there is no ambiguity regarding the nature of the fluctuation force field.
Therefore, we believe that with POM it is possible 
to derive a new semiclassical Langevin equation 
that is free from the problems mentioned above.

The purpose of this paper is to derive 
both quantum and semiclassical Langevin equations
within quantum field theory using POM and to compare the result so 
obtained with that obtained from IFM.
We first derive a quantum Langevin equation using POM.
In practice, 
it is difficult to numerically solve such a nonlinear operator equation.
For this reason, we derive a semiclassical Langevin equation from the
quantum equation employing a replacement procedure.
We apply POM and IFM to $\phi^4$ theory and 
derive Langevin equations for the soft modes projecting 
or integrating out the hard modes.
The method of deriving the semiclassical Langevin equation 
from the quantum Langevin equation is not unique.
Therefore, it is useful to compare 
the semiclassical Langevin equation obtained using IFM with 
the quantum Langevin equation obtained using POM 
and investigate the differences between their results beforehand.
In so doing, we show that the result given by POM is similar to 
that given by IFM with the linear harmonic approximation.
Furthermore, the differences between the fluctuation force fields of 
POM and IFM can be partially eliminated by utilizing the arbitrariness of 
the definition of the fluctuation force field in IFM.
Next, 
we discuss the derivation of the semiclassical Langevin equation in POM.

This paper is summarized as follows. 
In \S 2, POM is reviewed.
In \S 3, we apply POM to $\phi^4$ theory 
and derive a quantum Langevin equation for the soft modes.
In \S 4,
the derivation of the semiclassical Langevin equation in IFM is given.
This section follows the work of Greiner et al.\cite{ref:Grei-Mul}
The results of \S 3 and \S 4 are compared in \S 5.
In \S 6, we derive the semiclassical Langevin equation by replacing the  
relevant operators with c-numbers.
The result is compared with the semiclassical Langevin equation in IFM.
Conclusions and discussion are given in \S 7.

\section{Projection operator method}

The version of POM that we use in this paper 
has the following characteristics.
\cite{ref:H-S-S,ref:S-H,ref:U-S,ref:Koide-Maruyama}
First, it can be used in both Schr\"odinger and Heisenberg pictures; 
in other words, the master equation 
and the Langevin equation can be treated simultaneously.
Second, there is great freedom in the choice of the projection operators.
In fact, this projection operator method includes both 
the Mori and the Nakajima-Zwanzig methods.
\cite{ref:Mori,ref:Toku-Mori,ref:Naka-Zwa}
Third, the equations both with and without a time-convolution integral 
can be treated systematically.
Fourth, it can be used even when the Hamiltonian is explicitly time dependent.
In this paper, we derive the result in Heisenberg picture.
\cite{ref:Koide-Maruyama}

The starting point is the Heisenberg equation of motion,
\begin{eqnarray}
\frac{d}{dt}O(t) &=& i[H,O(t)] \nonumber \\
                 &=& iLO(t),
\end{eqnarray}
where $L$ is the Liouville operator.
We can rewrite the Heisenberg equation of motion by using
the projection operator, which has the following general properties:
\begin{eqnarray}
P^2 &=& P, \\
Q &=& 1-P, \\
PQ &=& QP = 0.
\end{eqnarray}
Then, the Heisenberg equation of motion can be rewritten as
\begin{eqnarray}
\frac{d}{dt}O(t)
&=& e^{iL(t-t_{0})}PiLO(t_{0})
+e^{iL(t-t_{0})}P\Sigma(t,t_{0})\frac{1}{1-\Sigma(t,t_{0})}iLO(t_{0}) 
\nonumber \\
&&+ Qe^{iLQ(t-t_{0})}\frac{1}{1-\Sigma(t,t_{0})}iLO(t_{0}), \nonumber \\
\label{eqn:genmitu-nt}
\end{eqnarray}
where
\begin{eqnarray}
     \Sigma(t,t_{0}) 
&\equiv& \int^{t}_{t_{0}}d{\tau}e^{-iL(t-\tau)}PiLQe^{iLQ(t-\tau)}.
\end{eqnarray}
This equation is called the time-convolutionless (TCL) equation because
it does not include time-convolution integral terms, as is shown below.
When the system has a dissipation effect,
it comes from the second term on the r.h.s. of the equation.
The third term represents the time variation 
in the space which is projected out.
Note that Q operates from the left in this term.
We regard this term as the fluctuation force term, 
as in the case of the Mori equation.
This equation is still equivalent to the Heisenberg equation of motion, 
and it is difficult to solve without approximation.
To make an approximation, we have to specify the projection operator.

We consider the case in which 
the total system can be divided into two parts, 
the system and the environment.
Then, the Hamiltonian is given by
\begin{eqnarray}
H &=& H_{S} + H_{E} + H_{I} \nonumber\\
  &=& H_{0} + H_{I},
\end{eqnarray}
where $H_{S}$ and $H_{E}$ are the unperturbed Hamiltonians
of the system and the environment, respectively,
and $H_{I}$ is the Hamiltonian that describes 
the interaction between the system and the environment.
The self-interactions of the system and the environment 
are also included in $H_{I}$.
We assume that the initial density matrix 
is given by the direct product of the system 
and the environment density matrices: 
\begin{eqnarray}
\rho = \rho_{S}\otimes\rho_{E}. \label{eqn:rho-dire}
\end{eqnarray}
We then define the projection operator $P$ as
\begin{eqnarray}
P O = {\rm Tr}_{E}[\rho_{E} O] = \langle O \rangle. \label{eqn:proj}
\end{eqnarray}
Here, ${\rm Tr}_{E}$ is the trace over only 
the environment degrees of freedom.

Using the above described projection operator,
we can rewrite the TCL equation as\cite{ref:Koide-Maruyama}
\begin{eqnarray}
 \frac{d}{dt}O(t)
&=&  e^{iL(t-t_{0})}PiLO(t_{0}) \nonumber \\
&&  -e^{iL(t-t_{0})}Pe^{-iL_{0}(t-t_{0})}{\cal C}(t,t_{0})Q
     \frac{1}{1+({\cal C}(t,t_{0})-1)Q}
     e^{iL_{0}(t-t_{0})}iLO(t_{0}) \nonumber \\
&&  + Qe^{iLQ(t-t_{0})}\frac{1}{1-\Sigma(t,t_{0})}iLO(t_{0}).
\label{eqn:genmitu-nt-2}
\end{eqnarray}
Here, the new function ${\cal C}(t,t_{0})$ is defined as
\begin{eqnarray}
\lefteqn{{\cal C}
  (t,t_{0})} && \nonumber \\
&=& e^{iL_{0}(t-t_{0})}e^{-iL(t-t_{0})} \nonumber \\
&=& 1+\sum^{\infty}_{n=1}(-i)^n \int^{t}_{t_{0}}dt_{1}\int^{t_{1}}_{t_{0}}dt_{2} \cdots \int^{t_{n-1}}_{t_{0}}dt_{n} \breve{L}_{I}(t_{1}-t_{0})\breve{L}_{I}(t_{2}-t_{0}) \cdots \breve{L}_{I}(t_{n}-t_{0}), \nonumber \\
\end{eqnarray}
where
\begin{eqnarray}
L_{0}O &=& [H_{0},O], \\
L_{I}O &=& [H_{I},O], \\
    \breve{L}_{I}(t-t_{0}) &\equiv&
e^{iL_{0}(t-t_{0})}L_{I}e^{-iL_{0}(t-t_{0})}.
\label{eqn:interaction}
\end{eqnarray}
The operator ${\cal C}(t,t_{0})$ is 
a time-ordered function of the Liouville operator.

For the second term on the r.h.s. of the equation,
we expand ${\cal C}(t,t_{0})Q/\{1+({\cal C}(t,t_{0})-1)Q\}$ up to
first order in $L_{I}$.
Then, we have an approximate representation of the TCL equation,
\begin{eqnarray}
  \frac{d}{dt}O(t) &=& e^{iL(t-t_{0})}Pe^{-iL_{0}(t-t_{0})}Pe^{iL_{0}(t-t_{0})}iLO(t_{0}) \nonumber \\
   && + e^{iL(t-t_{0})}Pe^{-iL_{0}(t-t_{0})}P\int^{t}_{t_{0}}ds e^{iL_{0}(s-t_{0})}iL_{I}e^{-iL_{0}(s-t_{0})}Qe^{iL_{0}(t-t_{0})}iLO(t_{0}) \nonumber \\
   && + Qe^{iLQ(t-t_{0})}\frac{1}{1-\Sigma(t,t_{0})}iLO(t_{0}). \label{eqn:2cd-tcl}
\end{eqnarray}
It can be seen that Eq. (\ref{eqn:2cd-tcl}) does not contain
a time-convolution integral,
from the form of the full time-evolution operator, $e^{iL(t-t_{0})}$,
which operates from the left in the second term on the r.h.s.
If this equation did contain a time-convolution integral,
the form of the full time-evolution operator must be $e^{iL(t-s)}$, where
$s$ is an integral variable.
This is the reason that the equation is called
the ``time-convolutionless equation''.

Now, we expand the third term.
It is not clear to what order we should expand this term.
Here, we apply the fluctuation-dissipation theorem of 
second kind (2nd f-d theorem) to the second and the third terms 
in order to fix the order of the expansion.
First, we rewrite the third term as
\begin{eqnarray}
\lefteqn{Qe^{iLQ(t-t_{0})}\frac{1}{1-\Sigma(t,t_{0})}iLO(t_{0})} 
&& \nonumber \\
&=& Qe^{iLQ(t-t_{0})}\frac{1}{1-Q\Sigma(t,t_{0})}iLO(t_{0}) \nonumber \\
&=&  Q{\cal D}(t,t_{0})Q \nonumber \\
&& \times \frac{1}{1+({\cal C}(t,t_{0})-1)Q+({\cal
D}(t,t_{0})-1)+({\cal C}(t,t_{0})-1)Q({\cal D}(t,t_{0})-1)}\nonumber
\\
&& \times e^{iL_{0}(t-t_{0})}iLO(t_{0}). 
\label{eqn:expa-flu}\nonumber \\
\end{eqnarray}
Here, the function ${\cal D}(t,t_{0})$ is defined as
\begin{eqnarray}
{\cal D}(t,t_{0})
&=& e^{iQLQ(t-t_{0})}e^{-iQL_{0}Q(t-t_{0})} \nonumber \\
&=& 1+\sum^{\infty}_{n=1} i^n
\int^{t}_{t_{0}}dt_{1}\int^{t_{1}}_{t_{0}}dt_{2} \cdots
\int^{t_{n-1}}_{t_{0}}dt_{n}
Q\breve{L}^{Q}_{I}(t_{n}-t_{0})Q\breve{L}^{Q}_{I}(t_{n-1}-t_{0})
\nonumber \\
&&\times \cdots Q\breve{L}^{Q}_{I}(t_{1}-t_{0}), \nonumber \\
\end{eqnarray}
where
\begin{eqnarray}
\breve{L}^{Q}_{I}(t-t_{0}) 
&\equiv& e^{iL_{0}(t-t_{0})}L_{I}Qe^{-iL_{0}(t-t_{0})}.
\end{eqnarray}
The function ${\cal D}(t,t_{0})$ is an anti-time-ordered function of 
the Liouville operator.
In the case of a linear Langevin equation, like the Mori equation, 
the 2nd f-d theorem, 
which relates the coefficient of the equation 
with the fluctuation force term, is exactly satisfied.
Therefore, we require it to be satisfied order by order 
in the perturbative expansion.
If we carry out the perturbative expansion up to zeroth order in 
$L_{I}$ for $Qe^{iLQ(t-t_{0})}\frac{1}{1-\Sigma(t,t_{0})}$ 
[that is, if we take ${\cal C}(t,t_{0})=1$ and ${\cal D}(t,t_{0})=1$ 
in Eq. (\ref{eqn:expa-flu})], 
the 2nd f-d theorem is satisfied in the case of
the linear Langevin equation, as shown in Appendix \ref{app:2ndfd}.
As a result, we have the approximate TCL equation
\begin{eqnarray}
  \frac{d}{dt}O(t) && = e^{iL(t-t_{0})}PiLO(t_{0}) \nonumber \\
   &&\hspace{-0.5cm} + e^{iL(t-t_{0})} \int^{t}_{t_{0}}ds
   e^{-iL_{0}(s-t_{0})}PiL_{I}e^{iL_{0}(s-t_{0})}Q
   iLO(t_{0}) \nonumber \\
   &&\hspace{-0.5cm} + Qe^{iL_{0}(t-t_{0})}iLO(t_{0}). \label{eqn:2cd-tcl-flu}
\end{eqnarray}
Here, we have used the additional 
condition $PL_{0}=L_{0}P$ for simplicity, because
in the following calculation, we consider
only the case in which this condition is satisfied.

In this section, we derived the TCL equation from the Heisenberg
equation of motion.
We can derive the TC equation that contains the
time-convolution integral similarly.
(See, for example, Appendix C of Ref.~\citen{ref:Koide-Maruyama}.)
In the present paper, we consider only the TCL equation because 
we wish to investigate the validity of the approximation employed in IFM in
order to eliminate the time-convolution integral.

We note here that the difference between the expressions 
``time-convolutionless'' and ``memoryless'' must be understood.
``Time-convolutionless'' does not mean that there is no memory effect. 
In fact, it has been confirmed that the TCL equation has a desirable
non-Markovian effect.\cite{ref:S-A,ref:Ahn}
Any memory effect that exists in the TC equation is also 
incorporated in the TCL equation through the higher-order terms, 
because these equations are equivalent.

\section{Quantum Langevin equation in POM}

In this section, we apply POM to $\phi^4$ theory.
The Lagrangian is given by
\begin{eqnarray}
{\cal L} = \frac{1}{2}\partial_{\mu}\hat{\phi}\partial^{\mu}\hat{\phi}
          -\frac{1}{2}m^2 \hat{\phi}^2-\frac{1}{4!}\lambda\hat{\phi}^4,
\end{eqnarray}
where the symbol $\hat{~}$ indicates an operator.
We set up the physical picture to be the same as that 
employed in Ref.~\citen{ref:Grei-Mul}.
In this paper, we treat the case $m^2 > 0$ and do not consider the symmetry
breaking, to avoid the complication involving the selection of the 
Fock space associated with degenerate vacua.
If we were to consider the case $m^2 < 0$, 
the system would have degenerate vacua and 
would undergo a phase transition to the symmetry broken phase.
In this case, the expectation value of the field operator $\hat{\phi}$ 
would play the role of an order parameter 
and would be regarded as a gross variable.
For this point, we consider only the $m^2 > 0$ case.
We may choose an initial conditions for which the order parameter 
is nonvanishing at $t=t_{0}$, due to its history before $t=t_{0}$.
The condensate will consist of soft modes.
We then introduce a cutoff $\Lambda_{I}$, which is the 
softest mode among the hard modes. 
That is, the hard modes are defined by $k \geq \Lambda_{I}$.
The largest value of $k$ among the modes contained 
in the condensate is assumed to be smaller than $\Lambda_{I}$.
We then impose another set of initial conditions in which the hard modes are
in thermal equilibrium with a high temperature $T$ at $t=t_{0}$.
In this situation, we regard the hard modes as irrelevant 
information and carry out a coarse-graining, as in the usual 
case of Brownian motion.
Then the hard modes become the origin of dissipation and fluctuation 
in the soft mode equation.

The field $\hat{\phi}({{\bf x}},t)$ can be divided 
into soft modes and hard modes as 
\begin{eqnarray}
\hat{\phi}({{\bf x}},t)
&=&  \int^{\Lambda_{I}}_{0}d^3 {\bf k}
    \frac{1}{\sqrt{2(2\pi)^3 \omega_{\bf k}}}
    [a_{\bf k}(t)e^{i{\bf kx}}+a^{\dagger}_{\bf k}(t)e^{-i{\bf kx}}]  \nonumber \\
&& + \int^{\Lambda}_{\Lambda_{I}} d^3 {\bf k}
    \frac{1}{\sqrt{2(2\pi)^3\omega_{\bf k}}}
    [a_{\bf k}(t)e^{i{\bf kx}}+a^{\dagger}_{\bf k}(t)e^{-i{\bf kx}}] \nonumber \\
&=& \hat{\phi}_{<}({\bf x},t) + \hat{\phi}_{>}({\bf x},t),
\end{eqnarray}
where $\Lambda$ is an ultraviolet cutoff 
and $\omega_{\bf k}=\sqrt{{\bf k}^2+m^2}$.
The field $\hat{\phi}_{<}({{\bf x}},t)$ contains only soft modes,
and the field $\hat{\phi}_{>}({{\bf x}},t)$ hard modes.
We consider the case in which the hard modes have already 
thermalized but the soft modes have not yet reached thermal 
equilibrium at $t=t_{0}$.
Therefore, the momentum cutoff $\Lambda_{I}$ 
is the lower limit of the momenta that constitute the thermalized 
hard modes.
Then, the hard modes play the role of a heat bath for the soft modes.
In this case, except for the vacuum polarization, which should be 
renormalized, 
the temperature $T$ acts as an effective cutoff 
for the hard modes, due to the Bose distribution function.
Therefore, to ensure that the hard modes  have a sufficiently large 
number of degrees of freedom, we must consider the situation that 
$\Lambda_{I}\ll T$.\cite{ref:Grei-Mul}

We are interested in only the slow evolution of the soft modes and 
hence have to coarse-grain the fast time dependence of the hard modes.
Consider the case of a small coupling constant.
Then, the time evolution of the soft modes and the hard modes is well 
approximated by that of the free field.
The time scale of the slowest soft mode is $\sim m^{-1}$, 
while that of the slowest hard mode is $\sim 1/\sqrt{\Lambda_{I}^2 + m^2}$.
To regard the hard modes as microscopic variables in relation to 
the soft modes, like a kind of random fluctuation, 
there must be a large difference 
between the typical time scales of the soft modes and the hard modes.
For this, the condition $m \ll \Lambda_{I}$ must be satisfied.
In short, we must impose the following condition for the cutoff 
$\Lambda_{I}$:
\begin{eqnarray}
m\ll \Lambda_{I}\ll T.
\end{eqnarray}
In this way, we finally obtain a semiclassical Langevin equation 
using POM in \S 6, in which the operator-valued fluctuation force terms 
are replaced with c-number stochastic variables.
Then, with the aforementioned initial conditions, 
the Langevin equation describes the time development of the 
condensate---actually, the decay or the dissolution of the condensate 
on a time scale longer than $1/\Lambda_{I}$.

The conjugate field $\hat{\Pi}({{\bf x}},t)$ is also divided as follows:
\begin{eqnarray}
\hat{\Pi}({{\bf x}},t)
&=&  -i\int^{\Lambda_{I}}_{0}d^3 {\bf k}
    \sqrt{\frac{\omega_{\bf k}}{2(2\pi)^3}}
    [a_{\bf k}(t)e^{i{\bf kx}}-a^{\dagger}_{\bf k}(t)e^{-i{\bf kx}}]  \nonumber \\
&& -i \int^{\Lambda}_{\Lambda_{I}} d^3 {\bf k}
    \sqrt{\frac{\omega_{\bf k}}{2(2\pi)^3}}
    [a_{\bf k}(t)e^{i{\bf kx}}-a^{\dagger}_{\bf k}(t)e^{-i{\bf kx}}] \nonumber \\
&=& \hat{\Pi}_{<}({\bf x},t) + \hat{\Pi}_{>}({\bf x},t).
\end{eqnarray}
The soft modes and the corresponding conjugate fields 
satisfy the commutation relations
\begin{eqnarray}
[ \hat{\phi}_{<}({\bf x},t),\hat{\phi}_{<}({\bf x}',t) ] &=& 0, \\
\protect[ \hat{\Pi}_{<}({\bf x},t),\hat{\Pi}_{<}({\bf x}',t) \protect] 
&=& 0, \\
\protect[\hat{\phi}_{<}({\bf x},t),\hat{\Pi}_{<}({\bf x}',t)\protect] &=& i\delta^{(3)}_{\Lambda_{I}}({{\bf x}}-{{\bf x}'}), \label{eqn:comutation}
\end{eqnarray}
where
\begin{eqnarray}
\delta^{(3)}_{\Lambda_{I}}({{\bf x}}-{{\bf x}'})
\equiv \frac{1}{(2\pi)^3}\int^{\Lambda_{I}}_{0} d^3 {\bf k}
e^{i{\bf k}({\bf x}-{\bf x}')} \label{eqn:l-delta}.
\end{eqnarray}
Here, the notation $\int^{\Lambda_{I}}_{0} d^3 {\bf k}$ means that
the integration is restricted to the region $0\leq|{\bf k}|\leq\Lambda_{I}$.
The function $\delta^{(3)}_{\Lambda_{I}}({{\bf x}}-{{\bf x}'})$
becomes the usual Dirac delta function
in the $\Lambda_{I} \rightarrow \infty$ limit.
Hereafter, we call it the ``cutoff delta function''.

The Hamiltonians of the system, the environment
and the interaction become
\begin{eqnarray}
H_{S} &=& \int d^{3}{\bf x} \frac{1}{2}[\hat{\Pi}^2_{<}({\bf x},t_{0})+(\nabla\hat{\phi}_{<}({\bf x},t_{0}))^2 +m^2\hat{\phi}^2_{<}({\bf x},t_{0})], \\
H_{E} &=& \int d^3 {\bf x} \frac{1}{2}[\hat{\Pi}^2_{>}({\bf x},t_{0})+(\nabla\hat{\phi}_{>}({\bf x},t_{0}))^2 +m^2\hat{\phi}^2_{>}({\bf x},t_{0})], \\
H_{I} &=& \int d^3 {\bf x}\frac{1}{4!}\lambda
  [\hat{\phi}^4_{<}({\bf x},t_{0})+4\hat{\phi}^3_{<}({\bf x},t_{0})\hat{\phi}_{>}({\bf x},t_{0})
  +6\hat{\phi}^2_{<}({\bf x},t_{0})\hat{\phi}^2_{>}({\bf x},t_{0}) \nonumber \\
&&+ 4\hat{\phi}_{<}({\bf x},t_{0})\hat{\phi}^3_{>}({\bf x},t_{0})
   +\hat{\phi}^4_{>}({\bf x},t_{0})].
\end{eqnarray}
Note that the division of $H$ into $H_{S}$, $H_{E}$ and $H_{I}$
is made at the initial time $t_{0}$.

The initial density matrix is given by the direct
product of the system and the environment density matrices
as in Eq. (\ref{eqn:rho-dire}).
We consider the case in which the environment is 
in thermal equilibrium with the Hamiltonian $H_{E}$, 
and thus 
\begin{eqnarray}
\rho_{E} &=& e^{-\beta H_{E}}/Z, \label{eqn:e-i-dmat}\\
Z &=& {\rm Tr}e^{-\beta H_{E}},
\end{eqnarray}
where $\beta=1/T$ is the inverse temperature.
Then, the projection operator is defined by Eq.~(\ref{eqn:proj}), 
with $\rho_{E}$ given by Eq.~(\ref{eqn:e-i-dmat}).

Substituting the above conditions into Eq.~(\ref{eqn:2cd-tcl}), 
we obtain the quantum Langevin equation.
Here, we summarize the approximations employed in our calculation.
We prepare the thermal equilibrium state as the initial environment, and 
expand ${\cal C}(t,t_{0})Q/\{1+({\cal C}(t,t_{0})-1)Q\}$ 
up to first order in $L_{I}$.
This approximation is made for the following reasons.
First, it is necessary to take into account at least first order in 
$L_{I}$ to have a dissipation effect.
Second, it was shown in a certain model 
that the system evolves toward the thermal equilibrium 
state at this level of approximation.\cite{ref:S-H-H-BL}
Furthermore, an approximation similar to that employed in this paper is 
employed in Ref.~\citen{ref:Xu-Grei}.
In that work, the nonlinear Langevin equation is solved numerically, 
and the temperature dependences of the order parameter and 
the effective masses of $\sigma$ and $\pi$ modes are determined.
These temperature dependences are the same as those 
exhibited when the system thermalizes 
with an initial environment temperature $T$.
Third, one of our main purposes is to compare our POM result 
with IFM result.\cite{ref:Grei-Mul}
The present approximation is sufficient for this purpose.

Setting the initial time as $t_{0}=0$ in Eq.~(\ref{eqn:2cd-tcl}), 
we obtain the quantum Langevin equation for the operator 
$\hat{\phi}_{<}({\bf x},t)$,
\begin{eqnarray}
\frac{d^2}{dt^2}\hat{\phi}_{<}({\bf x},t)
- \triangle\hat{\phi}_{<}({\bf x},t)+m^2\hat{\phi}_{<}({\bf x},t)
+ \frac{\lambda}{6}\int d^3 {\bf y} 
\delta^{(3)}_{{\Lambda}_{I}}({\bf x}-{\bf y})\hat{\phi}_{<}^3({\bf y},t) 
\hspace{-10cm}&& \nonumber \\
&& + \frac{\lambda}{2}\int d^3 {\bf y} \delta^{(3)}_{{\Lambda}_{I}}({\bf x}-{\bf y})
  \langle \hat{\phi}_{0>}^2 ({\bf y},t) \rangle \hat{\phi}_{<}({\bf y},t) \nonumber \\
&& +i\lambda^2\int^{t}_{0}ds \int d^3 {\bf y}_{1} d^3 {\bf y}_{2}
\delta^{(3)}_{{\Lambda}_{I}}({\bf x} - {\bf y}_{2})   \nonumber \\
&&\times \Bigg\{
\frac{1}{48}\langle [\hat{\phi}_{0>}^4 ({\bf y}_{1},s),\hat{\phi}_{0>}^2 ({\bf y}_{2},t)] \rangle \hat{\phi}_{<}({\bf y}_{2},t)
\nonumber \\
&&+ \frac{1}{36}\langle [\hat{\phi}_{0>}^3 ({\bf y}_{1},s),\hat{\phi}_{0>}^3 ({\bf y}_{2},t)] \rangle \hat{\varphi}_{<}({\bf y}_{2},s-t,t)
\nonumber \\
&& + \frac{1}{16}\langle [\hat{\phi}_{0>}^2 ({\bf y}_{1},s),\hat{\phi}_{0>}^2 ({\bf y}_{2},t)] \rangle [\hat{\varphi}_{<}^2 ({\bf y}_{1},s-t,t),\hat{\phi}_{<} ({\bf y}_{2},t)]_{+}
\nonumber \\
&& + \frac{1}{24} \langle[\hat{\phi}_{0>}({\bf y}_{1},s),\hat{\phi}_{0>}({\bf y}_{2},t)] \rangle
[\hat{\varphi}_{<}^3({\bf y}_{1},s-t,t),\hat{\phi}_{<}^2 ({\bf y}_{2},t)]
_{+}
\nonumber \\
&&+\frac{1}{8}\langle [\hat{\phi}_{0>} ({\bf
y}_{1},s),\hat{\phi}_{0>}({\bf y}_{2},t)]_{+} \rangle \nonumber \\
&&\times [\hat{\phi}_{0<}({\bf y}_{1},s-t),\hat{\phi}_{0<}({\bf y}_{2})]
[\hat{\varphi}^2_{<}({\bf y}_{1},s-t,t),\hat{\phi}_{<}({\bf y}_{2},t)]
_{+} \nonumber \\
&&+ \frac{1}{8}\langle [\hat{\phi}_{0>}^2 ({\bf
y}_{1},s),\hat{\phi}_{0>}^2 ({\bf y}_{2},t)-\langle \hat{\phi}_{0>}^2
({\bf y}_{2},t) \rangle ]_{+} \rangle \nonumber \\
&&\times [\hat{\varphi}_{<}({\bf y}_{1},s-t,t),\hat{\phi}_{<}({\bf y}_{2},t)]
\hat{\varphi}_{<}({\bf y}_{2},s-t,t)\Bigg\}
\nonumber \\
&&= \int d^3 {\bf y}\delta^{(3)} _{\Lambda_{I}}({\bf x}-{\bf y})
\{
\hat{f}_{1}({\bf y},t)
+\hat{f}_{2}({\bf y},t)\hat{\phi}_{0<} ({\bf y},t)
+\hat{f}_{3}\hat{\phi}_{0<}^2 ({\bf y},t)
\},  \label{eqn:p-phi4}
\end{eqnarray}
where
\begin{eqnarray}
\hat{f}_{1}({\bf x},t)
&=& \frac{\lambda}{6}\hat{\phi}_{0>}^3 ({\bf x},t), \\
\hat{f}_{2}({\bf x},t)
&=& \frac{\lambda}{2}(\hat{\phi}_{0>}^2 ({\bf x},t)-\langle \hat{\phi}_{0>}^2 ({\bf x},t) \rangle), \\
\hat{f}_{3}({\bf x},t)
&=& \frac{\lambda}{2}\hat{\phi}_{0>} ({\bf x},t).
\end{eqnarray}
Here, the fields with the subscript $0$ evolve as free fields, and
the newly introduced field $\hat{\varphi}_{<}({\bf x},s,t)$ is defined as
\begin{eqnarray}
\hat{\varphi}_{<}({\bf x},s,t)
&=&  \int^{\Lambda_{I}}_{0}d^3 {\bf k}
    \frac{1}{\sqrt{2(2\pi)^3 \omega_{\bf k}}}
    [a_{\bf k}(t) e^{-i\omega_{\bf k}s+i{\bf kx}} 
+a^{\dagger}_{\bf k}(t) e^{i\omega_{\bf k}s-i{\bf kx}}] \nonumber \\
&=& \frac{1}{(2\pi)^3}\int^{\Lambda_{I}}_{0}d^3 {\bf k}
    \{ \hat{\phi}_{<}({\bf k},t)\cos\omega_{\bf k}s + \frac{1}{\omega_{\bf k}}\hat{\Pi}_{<}({\bf k},t) \sin\omega_{\bf k}s \} e^{i{\bf kx}}, \nonumber \\
    \label{eqn:vp}
\end{eqnarray}
where
\begin{eqnarray}
\hat{\phi}_{<}({\bf k},t)
&=& \sqrt{\frac{(2\pi)^3}{2\omega_{\bf k}}}(a_{\bf k}(t) + a_{\bf -k}^{\dagger}(t)), \label{eqn:fou1}\\
\hat{\Pi}_{<}({\bf k},t)
&=& -i\sqrt{\frac{(2\pi)^3}{2\omega_{\bf k}}}(a_{\bf k}(t) - a_{\bf -k}^{\dagger}(t)). \label{eqn:fou2}
\end{eqnarray}
Furthermore, $[~]$ and $[~]_{+}$ in Eq.~(\ref{eqn:p-phi4})
represent the commutation and the anti-commutation relations, respectively.
The first four terms on the l.h.s. of this equation come from 
the first term on the r.h.s. of Eq.~(\ref{eqn:2cd-tcl-flu}).
The remaining terms on the l.h.s. come from the second term, and 
the terms on the r.h.s. come from the third term.
The diagrams that correspond to the interaction terms on the r.h.s. of
Eq. (\ref{eqn:p-phi4}) are shown in Fig. 1.
The diagrams (a) and (b) correspond to
the fourth and the fifth terms on the l.h.s. of the equation, respectively.
The diagrams (c) through (h) correspond to the first term through 
the sixth term within the brackets on the l.h.s. of the equation, respectively.

\begin{figure}[t]
\begin{center}\leavevmode
\epsfxsize=10cm
\epsfbox{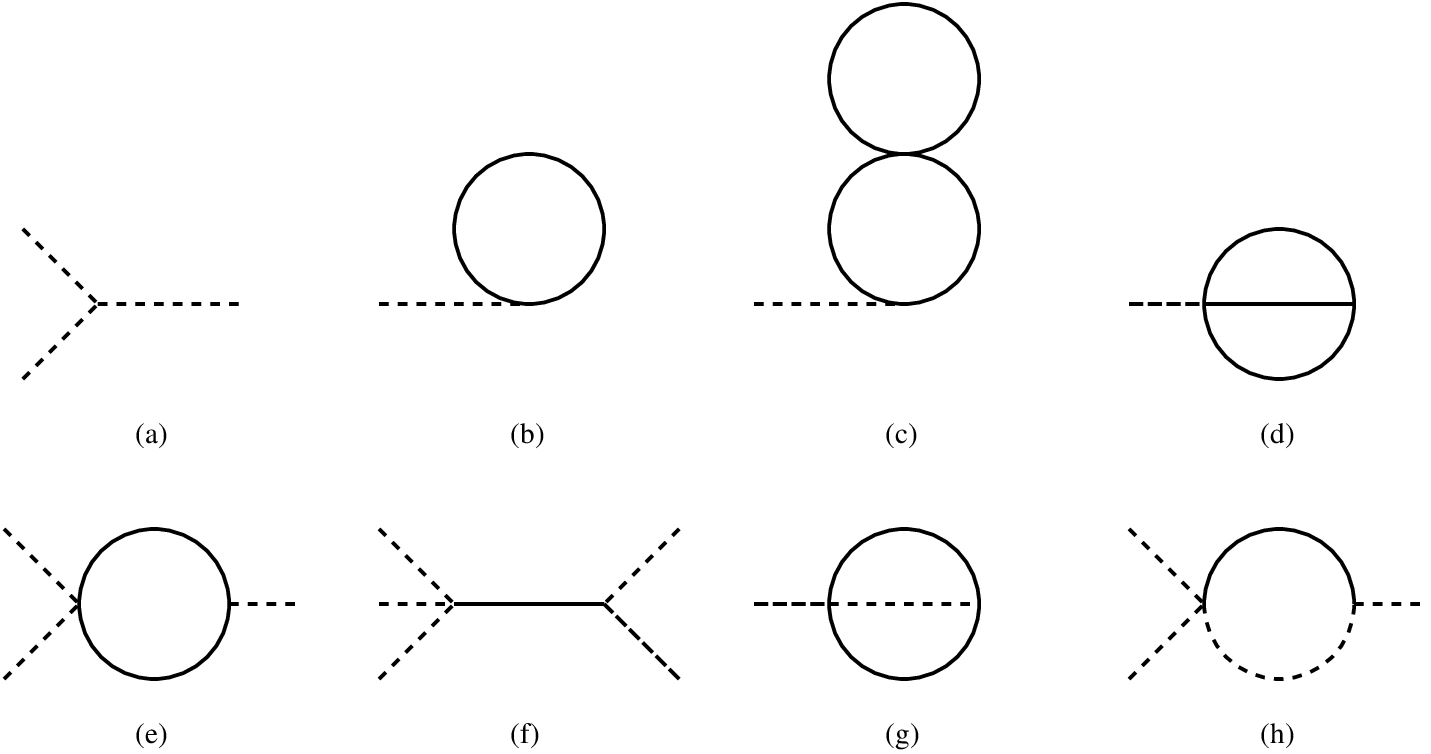}\\
Fig. 1 :Diagrams that contribute to 
the quantum Langevin equation in POM. 
The dashed lines represent the soft modes and the solid lines the hard modes.
\end{center} 
\end{figure}

Now, we call $\hat{f}_{i}$ the fluctuation force fields.
We examine the correlations of $\hat{f}_{i}$.
The expectation values of $\hat{f}_{i}$ can be calculated 
using the initial density matrix $\rho$.
However, it is also possible to replace $\rho$ with $\rho_{E}$, 
because $\hat{f}_{i}$ consists of only operators of the hard modes.
The first-order correlations are
\begin{eqnarray}
\langle \hat{f}_{1}({\bf x},t) \rangle =
\langle \hat{f}_{2}({\bf x},t) \rangle =
\langle \hat{f}_{3}({\bf x},t) \rangle = 0.
\end{eqnarray}
The second-order correlations are calculated as follows:
\begin{eqnarray}
\langle \hat{f}_{1}({\bf x}_{1},t_{1})\hat{f}_{1}({\bf x}_{2},t_{2}) \rangle
&=& \frac{i\lambda^2}{36} [6G_{\Lambda_{I}}({\bf x}_{1}-{\bf x}_{2},t_{1}-t_{2})^3+9 G_{\Lambda_{I}}({\bf x}_{1}-{\bf x}_{2},t_{1}-t_{2})G_{\Lambda_{I}}(0,0)^2], \nonumber \\
\label{eqn:flu-1}\\
\langle \hat{f}_{2}({\bf x}_{1},t_{1})\hat{f}_{2}({\bf x}_{2},t_{2}) \rangle
&=& -\frac{\lambda^2}{2}
 G_{\Lambda_{I}}({\bf x}_{1}-{\bf x}_{2},t_{1}-t_{2})^2, \label{eqn:flu-2}\\
\langle \hat{f}_{3}({\bf x}_{1},t_{1})\hat{f}_{3}({\bf x}_{2},t_{2}) \rangle
&=& -\frac{i\lambda^2}{4} G_{\Lambda_{I}}({\bf x}_{1}-{\bf x}_{2},t_{1}-t_{2}).
\label{eqn:flu-3}
\end{eqnarray}
The propagator on the r.h.s. of Eqs. (\ref{eqn:flu-1}), 
(\ref{eqn:flu-2}) and (\ref{eqn:flu-3}) is defined as 
\begin{eqnarray}
G_{\Lambda_{I}}({\bf x},t)
&=& i\int^{\Lambda}_{\Lambda_{I}} d^3 {\bf p}
\frac{1}{2(2\pi)^3 \omega_{\bf p}}
\{ (1+2n_{\omega_{\bf p}})\cos \omega_{\bf p}t -i \sin \omega_{\bf p}t \}e^{i{\bf p}{\bf x}}. \nonumber \\
\label{eqn:G><}
\end{eqnarray}
It has the property 
\begin{eqnarray}
G_{\Lambda_{I}}^{*}({\bf x},t) = -G_{\Lambda_{I}}(-{\bf x},-t).
\end{eqnarray}
This propagator includes only the hard modes.
The Bose distribution function $n_{\omega_{\bf p}}$ is given by 
\begin{eqnarray}
n_{\omega_{\bf p}} = \frac{1}{e^{\beta \omega_{\bf p}}-1}.
\end{eqnarray}

The correlation functions in Eqs.~(\ref{eqn:flu-1}), (\ref{eqn:flu-2}) 
and (\ref{eqn:flu-3}) are complex, 
and change form under the exchange 
${\bf x}_{1} \leftrightarrow {\bf x}_{2},t_{1} \leftrightarrow t_{2}$.
Now, we examine the real and the imaginary parts of the correlations.
The real parts of the second-order correlations are
\begin{eqnarray}
\langle [\hat{f}_{1}({\bf x}_{1},t_{1}),\hat{f}_{1}({\bf x}_{2},t_{2})]_{+} \rangle/2
&=& -\frac{\lambda^2}{12i}
[G_{\Lambda_{I}}({\bf x}_{1}-{\bf x}_{2},t_{1}-t_{2})^3 +G_{\Lambda_{I}}({\bf x}_{2}-{\bf x}_{1},t_{2}-t_{1})^3 ]
 \nonumber \\
&& +\frac{i\lambda^2}{8}G_{\Lambda_{I}}(0,0)^2
[G_{\Lambda_{I}}({\bf x}_{1}-{\bf x}_{2},t_{1}-t_{2}) +G_{\Lambda_{I}}({\bf x}_{2}-{\bf x}_{1},t_{2}-t_{1}) ]
, \nonumber \\
\label{eqn:f11} \\
\langle [\hat{f}_{2}({\bf x}_{1},t_{1}),\hat{f}_{2}({\bf x}_{2},t_{2})]_{+} \rangle/2
&=& -\frac{\lambda^2}{4}
[ G_{\Lambda_{I}}({\bf x}_{1}-{\bf x}_{2},t_{1}-t_{2})^2 + G_{\Lambda_{I}}({\bf x}_{2}-{\bf x}_{1},t_{2}-t_{1})^2 ]
 , \label{eqn:f22}\\
\langle [\hat{f}_{3}({\bf x}_{1},t_{1}),\hat{f}_{3}({\bf x}_{2},t_{2})]_{+} \rangle/2
&=& \frac{\lambda^2}{8i}
[ G_{\Lambda_{I}}({\bf x}_{1}-{\bf x}_{2},t_{1}-t_{2}) + G_{\Lambda_{I}}({\bf x}_{2}-{\bf x}_{1},t_{2}-t_{1}) ]. \label{eqn:f33} 
\end{eqnarray} 
We can see that these quantities are symmetric under the exchange of 
the arguments 
${\bf x}_{1} \leftrightarrow {\bf x}_{2},t_{1} \leftrightarrow t_{2}$.
These correlations play an important role in our investigation of 
the relation between the fluctuation force fields in POM and those in IFM.

The imaginary parts of the second-order correlations are
\begin{eqnarray}
\langle [\hat{f}_{1}({\bf x}_{1},t_{1}),\hat{f}_{1}({\bf x}_{2},t_{2})] \rangle/2
&=& \frac{\lambda^2}{72} \langle [\hat{\phi}_{0>}^3 ({\bf x}_{1},t_{1}),\hat{\phi}_{0>}^3 ({\bf x}_{2},t_{2})] \rangle, \label{eqn:2ndfd-1}\\
\langle [\hat{f}_{2}({\bf x}_{1},t_{1}),\hat{f}_{2}({\bf x}_{2},t_{2})] \rangle/2
&=& \frac{\lambda^2}{8} \langle [\hat{\phi}_{0>}^2 ({\bf x}_{1},t_{1}),\hat{\phi}_{0>}^2 ({\bf x}_{2},t_{2})] \rangle, \label{eqn:2ndfd-2}\\
\langle [\hat{f}_{3}({\bf x}_{1},t_{1}),\hat{f}_{3}({\bf x}_{2},t_{2})] \rangle/2
&=&
\frac{\lambda^2}{8} \langle [\hat{\phi}_{0>} ({\bf x}_{1},t_{1}),\hat{\phi}_{0>} ({\bf x}_{2},t_{2})] \rangle.
\label{eqn:2ndfd-3}
\end{eqnarray}
These quantities are related to the coefficients of the Langevin equation.
The Langevin equation can be expressed in terms of the above quantities as follows:
\begin{eqnarray}
&& \Box \hat{\phi}_{<}({\bf x},t) + m^2\hat{\phi}_{<}({\bf x},t)
+\frac{\lambda}{6}\int d^3 {\bf y} \delta^{(3)}_{{\Lambda}_{I}}({\bf x}-{\bf y})\hat{\phi}_{<}^3({\bf y},t)
+ \frac{\lambda}{2}\int d^3 {\bf y} \delta^{(3)}_{{\Lambda}_{I}}({\bf x}-{\bf y})
  \langle \hat{\phi}_{0>}^2 ({\bf y},t) \rangle \hat{\phi}_{<}({\bf y},t) \nonumber \\
&&\hspace{1.5cm} +i\int^{t}_{0}ds \int d^3 {\bf y}_{1} d^3 {\bf y}_{2}
\delta^{(3)}_{{\Lambda}_{I}}({\bf x} - {\bf y}_{2}) \nonumber \\
&&\hspace{1.5cm} \times\Bigg\{ \frac{\lambda^2}{48}\langle [\hat{\phi}_{0>}^4 ({\bf y}_{1},s),\hat{\phi}_{0>}^2 ({\bf y}_{2},t)] \rangle \hat{\phi}_{<}({\bf y}_{2},t) \nonumber \\
&&\hspace{1.5cm} + \langle [\hat{f}_{1}({\bf y}_{1},s),
\hat{f}_{1}({\bf y}_{2},t)] \rangle  \hat{\varphi}_{<}({\bf y}_{2},s-t,t) ] \nonumber \\
&&\hspace{1.5cm} + \frac{1}{4}\langle [\hat{f}_{2}({\bf y}_{1},s),\hat{f}_{2}({\bf y}_{2},t)] \rangle [\hat{\varphi}_{<}^2 ({\bf y}_{1},s-t,t),\hat{\phi}_{<} ({\bf y}_{2},t)]_{+}
\nonumber \\
&&\hspace{1.5cm} +\frac{1}{6}\langle [\hat{f}_{3}({\bf y}_{1},s),\hat{f}_{3}({\bf y}_{2},t)] \rangle
[\hat{\varphi}_{<}^3({\bf y}_{1},s-t,t),\hat{\phi}_{<}^2 ({\bf y}_{2},t)]_{+} \nonumber \\
&&+\frac{1}{8}\langle [\hat{\phi}_{0>} ({\bf y}_{1},s),\hat{\phi}_{0>}({\bf y}_{2},t)]_{+} \rangle
[\hat{\phi}_{0<}({\bf y}_{1},s-t),\hat{\phi}_{0<}({\bf y}_{2})]
[\hat{\varphi}^2_{<}({\bf y}_{1},s-t,t),\hat{\phi}_{<}({\bf y}_{2},t)]
_{+} \nonumber \\
&&+ \frac{1}{8}\langle [\hat{\phi}_{0>}^2 ({\bf
y}_{1},s),\hat{\phi}_{0>}^2 ({\bf y}_{2},t)-\langle \hat{\phi}_{0>}^2
({\bf y}_{2},t) \rangle ]_{+} \rangle 
[\hat{\varphi}_{<}({\bf y}_{1},s-t,t),\hat{\phi}_{<}({\bf y}_{2},t)]
\hat{\varphi}_{<}({\bf y}_{2},s-t,t)\Bigg\}
\nonumber \\
&&\hspace{1.5cm} = \int d^3 {\bf y}
\{
\hat{f}_{1}({\bf y},t)
+\hat{f}_{2}({\bf y},t)\hat{\phi}_{0<} ({\bf y},t)
+\hat{f}_{3}({\bf y},t)\hat{\phi}_{0<}^2 ({\bf y},t)
\}\delta^{(3)} _{\Lambda_{I}}({\bf x}-{\bf y}).  \label{eqn:cl-p-phi4}
\end{eqnarray}
In case of the nonlinear equation, it is in general not clear
what kind of relation should be satisfied between the coefficients 
of the Langevin equation and the fluctuation force field.
\footnote{
Shibata and Hashitsume studied the nonlinear Langevin equation in 
the quantal spin system interacting with a heat reservoir 
and derive a nonlinear fluctuation-dissipation theorem.
\cite{ref:S-H-flu}
}
In the present case, 
the fluctuation force fields 
are related to the coefficients of the Langevin equation that are similar to 
those given by the 2nd f-d theorem 
for the linear Langevin equation.\cite{ref:S-H,ref:S-H-flu}
For this reason, we regard these relations as constituting the 2nd f-d theorem 
for the nonlinear equation.
\footnote{It is worth noting that some terms of order $\lambda^2$ 
on the l.h.s.~of Eq.~(\ref{eqn:cl-p-phi4}) cannot be expressed in terms of 
$\hat{f}_{i}$.}

\section{Semiclassical Langevin equation in IFM}

In this section, we present the result of the application of IFM to 
$\phi^4$ theory.
It is essentially due to the work of Greiner et al.
\cite{ref:Grei-Mul}
(For a detailed derivation, see Appendix \ref{app:fv-qm}.)

The c-number field $\phi$ is divided into two parts again 
by employing the momentum cutoff $\Lambda_{I}$ as
\begin{eqnarray}
\phi({\bf x},t) &=& \frac{1}{(2\pi)^3}\int d^3 {\bf p} \{ \phi({\bf p},t)\theta(\Lambda_{I}-|{\bf p}|) + \phi({\bf p},t)\theta(|{\bf p}|-\Lambda_{I}) \}e^{i{\bf px}}\nonumber \\
 &=& \phi_{<}({\bf x},t)  + \phi_{>}({\bf x},t).
\end{eqnarray}
Here, the field $\phi_{<}({\bf x},t)$ is that of soft modes 
and $\phi_{>}({\bf x},t)$ that of hard modes.

The action of this system is
\begin{eqnarray}
S[\phi_{<},\phi_{>}]
&=& \int^{t}_{t_{0}}d^4 x
\left[
\frac{1}{2}\partial_{\mu}\phi_{<}\partial^{\mu}\phi_{<}
-\frac{1}{2}m^2\phi_{<}^2
+\frac{1}{2}\partial_{\mu}\phi_{>}\partial^{\mu}\phi_{>}
-\frac{1}{2}m^2\phi_{>}^2 \right. \nonumber \\
&& \left. -\frac{\lambda}{4!}
(\phi_{<}^4 + 4\phi_{<}^3\phi_{>} + 6\phi_{<}^2\phi_{>}^2 + 4\phi_{<}\phi_{>}^3 + \phi_{>}^4)
\right] \nonumber \\
&=& S_{\phi_{<}}[\phi_{<}] + S_{\phi_{>}}[\phi_{>}]
    + S_{int}[\phi_{<},\phi_{>}],
\end{eqnarray}
where
\begin{eqnarray}
S_{\phi_{<(>)}}[\phi_{<(>)}] &=& \int^{t}_{t_{0}}d^4 x \frac{1}{2} (\partial_{\mu}\phi_{<(>)}\partial^{\mu}\phi_{<(>)} -m^2\phi_{<(>)}^2 ), \\ S_{int}[\phi_{<},\phi_{>}] &=& -\int^{t}_{t_{0}}d^4 x \frac{\lambda}{4!} (\phi_{<}^4 + 4\phi_{<}^3\phi_{>} + 6\phi_{<}^2\phi_{>}^2 + 4\phi_{<}\phi_{>}^3 + \phi_{>}^4).
\end{eqnarray}
Now, as in case of POM, 
we assume that the initial density matrix of the total system
is given by the direct product of the system and the environment
density matrices:
\begin{eqnarray}
\rho(\phi_{<},\phi'_{<},\phi_{>},\phi'_{>};t_{0})
= \rho_{s}(\phi_{<},\phi'_{<};t_{0})\otimes\rho_{h}(\phi_{>},\phi'_{>};t_{0}).
\end{eqnarray}
Then, the influence functional is defined as
\begin{eqnarray}
e^{iS_{IF}[\phi_{<},\phi'_{<};t]}
&=& \int d\phi_{>f}d\phi_{>i}d\phi'_{>i}\int^{\phi_{>f}}_{\phi_{>i}}D\phi_{>}
  \int^{\phi_{>f}}_{\phi'_{>i}} D\phi'_{>} \nonumber \\
  &&\times \exp
  \{
  i( S_{\phi_{>}}[\phi_{>}]+S_{int}[\phi_{<},\phi_{>}]- S_{\phi_{>}}[\phi'_{>}]-S_{int}[\phi'_{<},\phi'_{>}] )
  \} \rho_{h}(\phi_{>i},\phi'_{>i};t_{0}), \nonumber \\
\end{eqnarray}
where the fields $\phi_{>i}$, $\phi'_{>i}$ and $\phi_{>f}$ correspond to 
the fields $\phi_{>}({\bf x},t_{0})$, $\phi'_{>}({\bf x},t_{0})$ 
and $\phi_{>}({\bf x},t)$, respectively.
$S_{IF}$ is called the ``influence action''.
Because it is difficult to calculate the influence action exactly,
we usually carry out a perturbative expansion.
In this calculation, we keep terms through the second order.
Then, we have
\begin{eqnarray}
{\rm Re}
S_{IF}[\phi_{<},\phi'_{<}]
&=& -\frac{\lambda}{2}\int^{t}_{t_{0}}d^4 x \phi_{\Delta}(x){\rm Im}[G^{\Lambda_{I}}_{++}(0)] \bar{\phi}(x) \nonumber \\
&& + \lambda^2\int^{t}_{t_{0}}d^4 x_{1} d^4 x_{2} \theta(t_{1}-t_{2})
\left\{
\frac{1}{2}\phi_{\Delta}(x_{1}){\rm Im}G^{\Lambda_{I}}_{++}(0){\rm Im}[G^{\Lambda_{I}}_{++}(x_{1}-x_{2})^2] \bar{\phi}(x_{1}) \right. \nonumber \\
&&\left. -\frac{1}{3}\phi_{\Delta}(x_{1}){\rm Re}[G^{\Lambda_{I}}_{++}(x_{1}-x_{2})^3]\bar{\phi}(x_{2}) \right. \nonumber \\
&& \left.
+\frac{1}{2}\phi_{\Delta}(x_{1})\bar{\phi}(x_{1}){\rm Im}[G^{\Lambda_{I}}_{++}(x_{1}-x_{2})^2] [\bar{\phi}^{2}(x_{2}) + \frac{1}{4}\phi_{\Delta}^{2}(x_{2})] \right. \nonumber \\
&&\left. +\frac{1}{6}\phi_{\Delta}(x_{1})[\bar{\phi^2}(x_{1})+\frac{1}{12}\phi^2_{\Delta}(x_{1})]{\rm Re}[G^{\Lambda_{I}}_{++}(x_{1}-x_{2})]\bar{\phi}(x_{2})[\bar{\phi}^2(x_{2}+\frac{3}{4}\phi_{\Delta}^{2}(x_{2}))]
\right\}, \nonumber \\
\\
{\rm Im}
S_{IF}[\phi_{<},\phi'_{<}]
&=& \lambda^2 \int^{t}_{t_{0}}d^4 x_{1} d^4 x_{2}
\left\{
-\frac{1}{12}\phi_{\Delta}(x_{1}){\rm Im}[G^{\Lambda_{I}}_{++}(x_{1}-x_{2})^3]\phi_{\Delta}(x_{2}) \right. \nonumber \\
&& \left. -\frac{1}{4}\phi_{\Delta}(x_{1})\bar{\phi}(x_{1})
{\rm Re}
[G^{\Lambda_{I}}_{++}(x_{1}-x_{2})^2]
\phi_{\Delta}(x_{2})\bar{\phi}(x_{2}) \right.
\nonumber \\
&& \left. +\frac{1}{8}\phi_{\Delta}(x_{1})[\bar{\phi}^2(x_{1})+\frac{1}{12}\phi^2_{\Delta}(x_{1})]{\rm Im}[G^{\Lambda_{I}}_{++}(x_{1}-x_{2})]\phi_{\Delta}(x_{2})[\bar{\phi}^2(x_{2}) + \frac{1}{12}\phi_{\Delta}^2(x_{2})]
\right\}, \nonumber \\
\label{eqn:IA-IM}
\end{eqnarray}
where
\begin{eqnarray}
\bar{\phi}(x) = \frac{1}{2}(\phi_{<}(x) + \phi'_{<}(x)),
~~~~\phi_{\Delta}(x) = \phi_{<}(x) - \phi'_{<}(x).
\end{eqnarray}
Here, we have introduced the propagator
\begin{eqnarray}
G^{\Lambda_{I}}_{++}(x_{1}-x_{2})
&=& i\langle T[\phi_{>}(x_{1})\phi_{>}(x_{2})] \rangle \nonumber \\
&=& \theta(t_{1}-t_{2})i\langle \phi_{>}(x_{1})\phi_{>}(x_{2}) \rangle
   + \theta(t_{2}-t_{1})i\langle \phi_{>}(x_{2})\phi_{>}(x_{1}) \rangle \nonumber \\
&=& \theta(t_{1}-t_{2})G_{\Lambda_{I}}({\bf x}_{1}-{\bf x}_{2},t_{1}-t_{2})
   + \theta(t_{2}-t_{1})G_{\Lambda_{I}}({\bf x}_{2}-{\bf x}_{1},t_{2}-t_{1}). \nonumber \\
 \label{eqn:f-v-green}
\end{eqnarray}
Furthermore, we introduce the following notation for simplicity:
\begin{eqnarray}
{\rm Re}
[G_{++}^{\Lambda_{I}}(x_{1}-x_{2})^n] &=&
[\theta(t_{1}-t_{2})+(-1)^n \theta(t_{1}-t_{2})] \nonumber \\
&&\times \frac{1}{2}[G_{\Lambda_{I}}({\bf x}_{1}-{\bf x}_{2},t_{1}-t_{2})^n +(-1)^n G_{\Lambda_{I}}({\bf x}_{2}-{\bf x}_{1},t_{2}-t_{1})^n ], \nonumber \\ \\ {\rm Im}[G_{++}^{\Lambda_{I}}(x_{1}-x_{2})^n] &=& [\theta(t_{1}-t_{2})-(-1)^n \theta(t_{1}-t_{2})] \nonumber \\ &&\times \frac{1}{2i}[G_{\Lambda_{I}}({\bf x}_{1}-{\bf x}_{2},t_{1}-t_{2})^n -(-1)^n G_{\Lambda_{I}}({\bf x}_{2}-{\bf x}_{1},t_{2}-t_{1})^n ]. \nonumber \\
\end{eqnarray}

Note that the influence action $S_{IF}$ is complex in general.
Our next task is to define a new action $\tilde{S}_{IF}$ 
which is purely real.
It is called the ``stochastic influence action'' and is defined as
\begin{eqnarray}
e^{iS_{IF}[\phi_{<},\phi'_{<}]} =\int D\xi_{1} D\xi_{2} D\xi_{3} P_{1}[\xi_{1}]P_{2}[\xi_{2}]P_{3}[\xi_{3}]e^{i\tilde{S}_{IF}[\phi_{<},\phi'_{<};\xi_{i}]}. \label{eqn:sif}
\end{eqnarray}
Here, the stochastic weights $P_{i}$ are given by
\begin{eqnarray}
P_{i}[\xi_{i}] &=& N_{i}\exp \left\{ \int^{t}_{t_{0}}d^4 x_{1}d^4 x_{2}\xi_{i}(x_{1})I^{-1}_{i}(x_{1}-x_{2})\xi_{i}(x_{2}) \right \}, \end{eqnarray}
where
\begin{eqnarray}
I^{-1}_{1}(x_{1}-x_{2})
&=& \frac{3}{\lambda^2}[{\rm Im}G^{\Lambda_{I}}_{++}(x_{1}-x_{2})^3]^{-1}, \label{eqn:i1}\\
I^{-1}_{2}(x_{1}-x_{2})
&=& \frac{1}{\lambda^2}[{\rm Re}G^{\Lambda_{I}}_{++}(x_{1}-x_{2})^2]^{-1}\label{eqn:i2}, \\
I^{-1}_{3}(x_{1}-x_{2})
&=& -\frac{2}{\lambda^2}[{\rm Im}G^{\Lambda_{I}}_{++}(x_{1}-x_{2})]^{-1}\label{eqn:i3}.
\end{eqnarray}
Here, $N_{i}$ is a normalization constant.
The auxiliary fields $\xi_{1}$, $\xi_{2}$ and $\xi_{3}$ have been introduced 
to eliminate the imaginary part of the influence action.
Then, the stochastic influence action $\tilde{S}_{IF}$ is given by
\begin{eqnarray}
\tilde{S}_{IF}[\phi_{<},\phi'_{<};\xi_{i}] = {\rm Re}[S_{IF}[\phi_{<},\phi'_{<}]]
+\sum^{3}_{N=1}\frac{1}{N}\int^{t}_{t_{0}}d^4 x
(\phi^{N}_{<}(x)-\phi'^{N}_{<}(x))
\xi_{N}(x).\nonumber \\
 \label{eqn:til-sif}
\end{eqnarray}
To prove Eq. (\ref{eqn:til-sif}), we have to assume the relation 
\begin{eqnarray}
e^{-(1/2)\chi^{T}A\chi} = [det(2\pi A)]^{-1/2}\int D\xi \exp
\left( -\frac{1}{2}\xi^{T}A^{-1}\xi +i\chi^{T}\xi \right). \label{eqn:phi-gaussf}
\end{eqnarray}
(See Appendix C for details.)

Finally, the stochastic effective action is defined as
\begin{eqnarray}
\tilde{S}_{eff}[\bar{\phi},\phi_{\Delta},\xi_{i}] \equiv S[\phi_{<}]- S[\phi_{<}']+\tilde{S}_{IF}[\phi_{<},\phi_{<}';\xi_{i}]. \label{eqn:sto-eac}
\end{eqnarray}
Note that this action is real.
If the soft modes behave quasiclassically,
we can apply the variational relation 
\begin{eqnarray}
\left. \frac{\delta \tilde{S}_{eff}[\bar{\phi},\phi_{\Delta},\xi_{i}]}{\delta \phi_{\Delta}} \right|_{\phi_{\Delta}=0} = 0
\end{eqnarray}
to the stochastic effective action 
to obtain a semiclassical Langevin equation.
In order to carry out this functional derivative,
we must define a functional derivative for the soft modes.
Greiner et al. used the usual definition,
\begin{eqnarray}
\frac{\delta\phi_{<}(x')}{\delta\phi_{<}(x)} = \delta^{(4)}(x'-x). \label{eqn:g-m-henbun}
\end{eqnarray}
Thus we obtain the semiclassical Langevin equation
\begin{eqnarray}
&&\Box \phi_{<}({\bf x},t) + m^{2}\phi_{<}({\bf x},t)
 + \left.
      \sum_{i=a,\cdots,f}(-1)\frac{\delta{\rm Re}[S^{(i)}_{IF}[\phi_{<},\phi'_{<}]]}{\delta \phi_{\Delta}} \right|_{\phi_{\Delta=0}} \nonumber \\
&&\hspace*{4cm} =[\xi_{1}({\bf x},t)+\phi_{<}({\bf x},t)\xi_{2}({\bf x},t)+\phi^2_{<}({\bf x},t) \xi_{3}({\bf x},t)], \nonumber \\
\label{eqn:f-v-phi4}
\end{eqnarray}
where
\begin{eqnarray}
\left.
-\frac{\delta{\rm Re}[S^{(a)}_{IF}]}
{\delta \phi_{\Delta}(x)}
\right|_{\phi_{\Delta=0}}
&=& \frac{\lambda}{6}\phi_{<}^3({\bf x},t), \\
\left.
-\frac{\delta{\rm Re}[S^{(b)}_{IF}]}{\delta \phi_{\Delta}(x)}\right|_{\phi_{\Delta=0}}
&=& -\frac{i\lambda}{2}G_{\Lambda_{I}}(0,0) \phi_{<}({\bf x},t), \label{eqn:sifa}\\
\left.
-\frac{\delta{\rm Re}[S^{(c)}_{IF}]}{\delta \phi_{\Delta}(x)}\right|_{\phi_{\Delta=0}}
&=& \int^{t-t_{0}}_{0}d \tau \int d^3 {\bf x}'\frac{\lambda^2}{4}
G_{\Lambda_{I}}(0,0)[ G_{\Lambda_{I}}({\bf x}-{\bf x}',\tau)^2 - G_{\Lambda_{I}}({\bf x}'-{\bf x},-\tau)^2 ]\phi_{<}({\bf x},t), \nonumber \\
\label{eqn:sifb}\\
\left.
-\frac{\delta{\rm Re}[S^{(d)}_{IF}]}{\delta \phi_{\Delta}(x)}\right|_{\phi_{\Delta=0}}
&=& \int^{t-t_{0}}_{0}d\tau \int d^3 {\bf x}'\frac{\lambda^2}{6}
[G_{\Lambda_{I}}({\bf x}-{\bf x}',\tau)^3-G_{\Lambda_{I}}({\bf x}'-{\bf x},-\tau)^3]
\phi_{<}({\bf x}',t-\tau), \label{eqn:sifc}\\
\left.
-\frac{\delta{\rm Re}[S^{(e)}_{IF}]}{\delta \phi_{\Delta}(x)}\right|_{\phi_{\Delta=0}}
&=& \phi_{<}(x)\int^{t-t_{0}}_{0}d\tau \int d^3 {\bf x}'\frac{i\lambda^2}{4}
[G_{\Lambda_{I}}({\bf x}-{\bf x}',\tau)^2-G_{\Lambda_{I}}({\bf x}'-{\bf x},-\tau)^2]
\phi^2_{<}({\bf x}',t-\tau),  \nonumber \\
\label{eqn:sifd}\\
\left.
-\frac{\delta{\rm Re}[S^{(f)}_{IF}]}{\delta \phi_{\Delta}(x)}\right|_{\phi_{\Delta=0}}
&=& \phi^2_{<}(x)\int^{t-t_{0}}_{0}d\tau \int d^3 {\bf x}'\frac{-\lambda^2}{12}
[G_{\Lambda_{I}}({\bf x}-{\bf x}',\tau)-G_{\Lambda_{I}}({\bf x}'-{\bf x},-\tau)]
\phi^3_{<}({\bf x}',t-\tau). \nonumber \\
\label{eqn:sife}
\end{eqnarray}
Here, the superscripts (a),(b),$\cdots$ correspond, respectively, 
to the diagrams (a),(b),$\cdots$ in Fig. 2.
This classical treatment of the soft modes is justified 
in weakly coupled bosonic quantum fields 
at high temperature.
(See Refs. \citen{ref:Grei-Mul}, \citen{ref:Rish} 
and \citen{ref:Xu-Grei} for details.)

\begin{figure}[t]
\begin{center}\leavevmode
\epsfxsize=10cm
\epsfbox{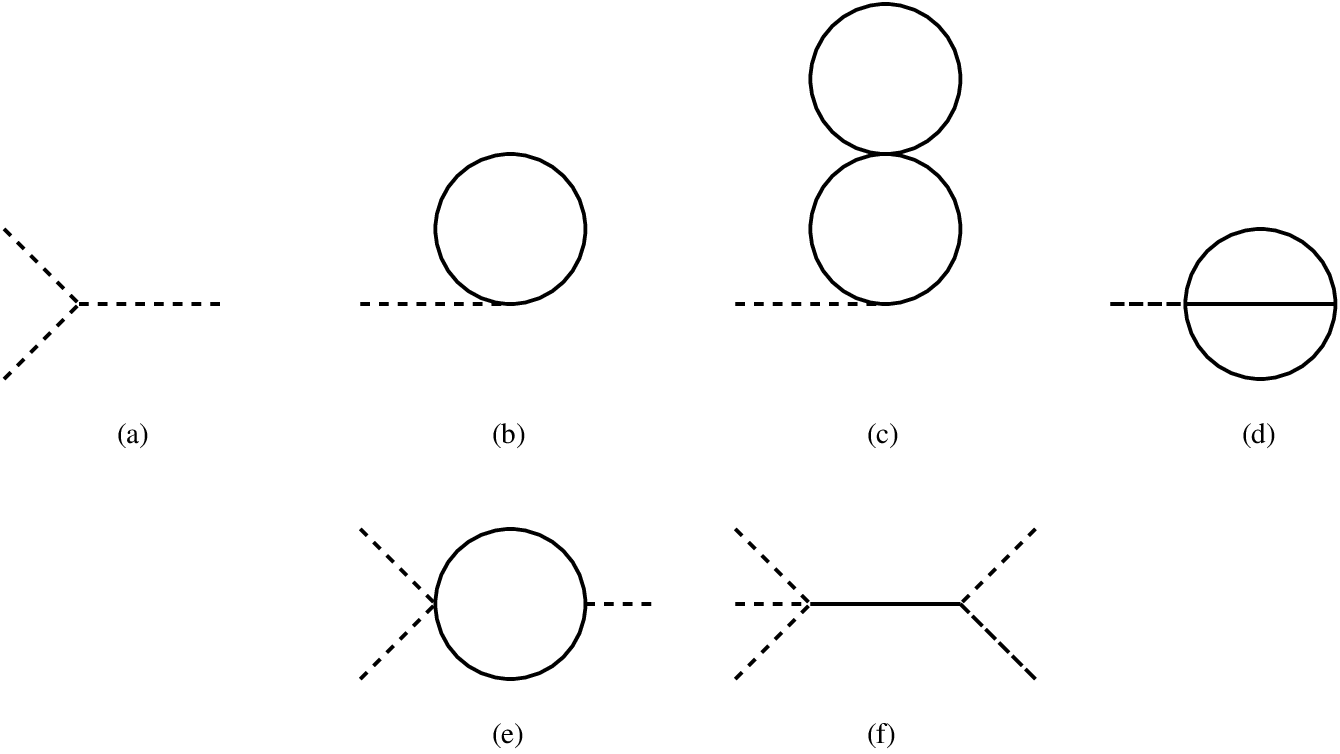}\\
Fig. 2 :Diagrams that contribute to the semiclassical Langevin equation 
in IFM. 
The dashed lines represent the soft modes and the solid lines the hard modes.
\end{center} 
\end{figure}

In IFM, the derived semiclassical equation in general possesses a 
time-convolution integral.
In the present case, it is contained 
in Eqs. (\ref{eqn:sifc}), (\ref{eqn:sifd}) and (\ref{eqn:sife}).
To remove such an integral, Greiner et al. introduced
an approximation called
the ``linear harmonic approximation''.
To explain this approximation, 
we consider the equation 
\begin{eqnarray}
\frac{d}{dt}a(t) = -i\omega a(t) + g\int^{t}_{0}ds F(s) a(t-s) 
\label{eqn:exa-tc}
\end{eqnarray}
for some variable $a(t)$, 
where $\omega$ is a frequency, $g$ is a coupling constant 
and $F(s)$ is an appropriate function of $s$.
This equation has a time-convolution integral 
in the second term on the r.h.s.
The quasi-instantaneous approximation is often used 
to remove such an integral.
In this approximation, one makes the replacement
\begin{eqnarray}
a(t-s) \longrightarrow a(t) -s \frac{da(t)}{dt}.
\end{eqnarray}
This approximation is justified when the time dependence of $a(t)$ 
is sufficiently weak.\cite{ref:Rau-Mul}
However,
if we use this approximation in $\phi^4$ theory, the dissipation 
effect is lost.
To retain the dissipation effect, an additional approximation is needed.
For example, one may use a propagator with an explicit width instead of the
propagator defined in Eq.~(\ref{eqn:f-v-green}). 
\cite{ref:Morikawa,ref:Glei-Ramo}

The linear harmonic approximation recently introduced by Greiner et al. \cite{ref:Grei-Mul} implies the replacement
\begin{eqnarray}
a(t-s) \longrightarrow  a(t)e^{i\omega s}.
\end{eqnarray}
Here, the dependence of $a(t-s)$ on the integration variable $s$
is replaced with that of the free vibration.
It is obvious that the time-convolution integral in Eq.~(\ref{eqn:exa-tc}) 
is eliminated through this replacement.
In this approximation,
we can retain the dissipation effect without imposing any further approximation
in $\phi^4$ theory.
However, the physical meaning of the approximation is not clear.

In $\phi^4$ theory, the linear harmonic approximation implies 
the replacement
\begin{eqnarray}
\phi_{<}({\bf k},t-\tau) \longrightarrow \phi_{<}({\bf k},t)\cos \omega_{\bf k}\tau
- \dot{\phi}_{<}({\bf k},t)\frac{1}{\omega_{\bf k}}\sin \omega_{\bf k}\tau.
\end{eqnarray}
With this replacement, the three terms defined in Eqs.~(\ref{eqn:sifc}), 
(\ref{eqn:sifd}) and (\ref{eqn:sife}) 
can be rewritten as follows:
\begin{eqnarray}
\left.
-\frac{\delta{\rm Re}[S^{(d)}_{IF}]}{\delta \phi_{\Delta}(x)}\right|_{\phi_{\Delta=0}} &=& \int^{t-t_{0}}_{0}d\tau \int d^3 {\bf x}'\frac{\lambda^2}{6}
[G_{\Lambda_{I}}({\bf x}',\tau)^3-G_{\Lambda_{I}}(-{\bf x}',-\tau)^3]
\varphi_{<}({\bf x}-{\bf x}',-\tau,t), \label{eqn:lh-sifc}\\
\left.
-\frac{\delta{\rm Re}[S^{(e)}_{IF}]}{\delta \phi_{\Delta}(x)}\right|_{\phi_{\Delta=0}} &=& \phi_{<}(x)\int^{t-t_{0}}_{0}d\tau \int d^3 {\bf x}'\frac{i\lambda^2}{4}
[G_{\Lambda_{I}}({\bf x}',\tau)^2-G_{\Lambda_{I}}(-{\bf x}',-\tau)^2]
\varphi_{<}({\bf x}-{\bf x}',-\tau,t)^2, \nonumber \\
\label{eqn:lh-sifd}\\
\left.
-\frac{\delta{\rm Re}[S^{(f)}_{IF}]}{\delta \phi_{\Delta}(x)}\right|_{\phi_{\Delta=0}} &=& \phi^2_{<}(x)\int^{t-t_{0}}_{0}d\tau \int d^3 {\bf x}'\frac{-\lambda^2}{12}
[G_{\Lambda_{I}}({\bf x}',\tau)-G_{\Lambda_{I}}(-{\bf x}',-\tau)]
\varphi_{<}({\bf x}-{\bf x}',-\tau,t)^3. \nonumber \\
\label{eqn:lh-sife}
\end{eqnarray}
Here, we have introduced the function [cf. Eq.~(\ref{eqn:vp})]
\begin{eqnarray}
\varphi_{<}({\bf x},\tau,t)
= \frac{1}{(2\pi)^3}\int^{\Lambda_{I}}_{0}d^3 {\bf k}
  (\phi_{<}({\bf k},t)\cos \omega_{\bf k}\tau + \dot{\phi}_{<}({\bf k},t) \frac{1}{\omega_{\bf k}}\sin \omega_{\bf k}\tau )
   e^{i{\bf kx}}. \nonumber \\
\end{eqnarray}
Finally, the semiclassical Langevin equation without 
a time-convolution integral is obtained
if we use the above expressions instead of 
Eqs.~(\ref{eqn:sifc}), (\ref{eqn:sifd}) and (\ref{eqn:sife}).

Now, we examine the r.h.s. of Eq.~(\ref{eqn:f-v-phi4}).
In the IFM calculation, the auxiliary fields $\xi_{i}$ are interpreted
as the fluctuation force field.\cite{ref:Sch,ref:Cal-Legg}
Their expectation values are 
calculated by using the stochastic weights $P_{i}$.
Then, the first order correlations are
\begin{eqnarray}
\langle \xi_{1}({\bf x},t) \rangle_{P}
= \langle \xi_{2}({\bf x},t) \rangle_{P}
= \langle \xi_{3}({\bf x},t) \rangle_{P}
= 0,
\end{eqnarray}
where
\begin{eqnarray}
\langle \xi_{i}({\bf x},t) \rangle_{P}
=\int D\xi_{i} P_{i}[\xi_{i}]\xi_{i}({\bf x},t).
\end{eqnarray}
To know the characteristics of the correlations, 
it is necessary to calculate the correlations up to at least second-order.
We have
\begin{eqnarray}
\langle \xi_{1}({\bf x}_{1},t_{1})\xi_{1}({\bf x}_{2},t_{2}) \rangle_{P}
&=& -\frac{\lambda^2}{12i}[G_{\Lambda_{I}}({\bf x}_{1}-{\bf x}_{2},t_{1}-t_{2})^3 + G_{\Lambda_{I}}({\bf x}_{2}-{\bf x}_{1},t_{2}-t_{1})^3], \nonumber \\
 \label{eqn:xi11}\\
\langle \xi_{2}({\bf x}_{1},t_{1})\xi_{2}({\bf x}_{2},t_{2}) \rangle_{P}
&=& -\frac{\lambda^2}{4}[G_{\Lambda_{I}}({\bf x}_{1}-{\bf x}_{2},t_{1}-t_{2})^2 + G_{\Lambda_{I}}({\bf x}_{2}-{\bf x}_{1},t_{2}-t_{1})^2], \nonumber \\
 \label{eqn:xi22}\\
\langle \xi_{3}({\bf x}_{1},t_{1})\xi_{3}({\bf x}_{2},t_{2}) \rangle_{P}
&=& \frac{\lambda^2}{8i}[G_{\Lambda_{I}}({\bf x}_{1}-{\bf x}_{2},t_{1}-t_{2}) + G_{\Lambda_{I}}({\bf x}_{2}-{\bf x}_{1},t_{2}-t_{1})]  \label{eqn:xi33}, \\
\langle \xi_{i}({\bf x}_{1},t_{1})\xi_{j}({\bf x}_{2},t_{2}) \rangle_{P}
&=& 0.   ~~~~~~~~~~~~~~~({\rm for~i \neq j})
\end{eqnarray}
It is noteworthy that 
there is no correlation between $\xi_{i}$ and $\xi_{j}$ for $i \neq j$.

In \S 3, we have pointed out that, in POM, 
the 2nd f-d theorem is satisfied 
for the coefficients and the fluctuation force fields. 
In IFM, it is said that there is a relation called 
the ``generalized fluctuation-dissipation theorem''.
\cite{ref:Hu-Paz-Zha,ref:Hu-Paz-Zha2}
To confirm the validity of this relation, 
it is convenient to introduce the following
functions ${\cal M}^{(i)}({\bf k},\omega)$:
\begin{eqnarray}
\lefteqn{\left. -\frac{\delta{\rm Re}[S^{(d)}_{IF}]}{\delta \phi_{\Delta}(x)}\right|_{\phi_{\Delta=0}} } && \nonumber \\
&=& \int \frac{d^3 {\bf k}}{(2\pi)^3}e^{\i{\bf kx}}\theta(\Lambda_{I}-|{\bf k}|)
\int^{t-t_{0}}_{0}d\tau \varphi_{<}({\bf k},-\tau,t) \int \frac{d\omega}{2\pi}
e^{-i\omega\tau}{\cal M}^{(1)}({\bf k},\omega), \nonumber \\
\label{eqn:m-11}\\
\lefteqn{\left.-\frac{\delta{\rm Re}[S^{(e)}_{IF}]}{\delta \phi_{\Delta}(x)}\right|_{\phi_{\Delta=0}} } && \nonumber \\
&=& \phi_{<}(x)\int \frac{d^3 {\bf k}}{(2\pi)^3}e^{\i{\bf kx}} \int^{\Lambda_{I}}_{0} \frac{d^3 {\bf k}_{1}}{(2\pi)^3} \theta(\Lambda_{I}-|{\bf k}-{\bf k}_{1}|) \nonumber \\
&& \times \int^{t-t_{0}}_{0}d\tau
\varphi_{<}({\bf k}-{\bf k}_{1},-\tau,t)\varphi_{<}({\bf k}_{1},-\tau,t)
 \int \frac{d\omega}{2\pi}
e^{-i\omega\tau}{\cal M}^{(2)}({\bf k},\omega),  \nonumber \\
\label{eqn:m-22}\\
\lefteqn{\left.-\frac{\delta{\rm Re}[S^{(f)}_{IF}]}{\delta \phi_{\Delta}(x)}\right|_{\phi_{\Delta=0}} } && \nonumber \\
&=& \phi_{<}^{2}(x)\int \frac{d^3 {\bf k}}{(2\pi)^3}e^{\i{\bf kx}}
\int^{\Lambda_{I}}_{0} \frac{d^3 {\bf k}_{1}d^3 {\bf k}_{2}}{(2\pi)^6}
\theta(\Lambda_{I}-|{\bf k}-{\bf k}_{1}-{\bf k}_{2}|) \nonumber \\
&& \times \int^{t-t_{0}}_{0}d\tau
\varphi_{<}({\bf k}-{\bf k}_{1}-{\bf k}_{2},-\tau,t)
\varphi_{<}({\bf k}_{1},-\tau,t)
\int \frac{d\omega}{2\pi}
e^{-i\omega\tau}{\cal M}^{(3)}({\bf k},\omega). \nonumber \\
\label{eqn:m-33}
\end{eqnarray}
Here, $\varphi_{<}({\bf k},\tau,t)$ is the Fourier transform of 
$\varphi_{<}({\bf x},\tau,t)$.
Furthermore,
the Fourier transform of the function $I_{i}(x)$, 
which is the inverse of the matrix 
$I^{-1}_{i}(x)$ given by Eq.~(\ref{eqn:i1}), (\ref{eqn:i2})
and (\ref{eqn:i3}), is defined as 
\begin{eqnarray}
I_{i}(x)
= \int \frac{d\omega d^3 {\bf k}}{(2\pi)^4} e^{i{\bf kx}}e^{-i\omega t}I_{i}({\bf k},\omega).
\end{eqnarray}
Then, we obtain the relation
\begin{eqnarray}
I_{i}({\bf k},\omega) = \frac{i}{\omega}
K^{(i)}(\omega) {\cal M}^{(i)}({\bf k},\omega),
\end{eqnarray}
where
\begin{eqnarray}
K^{(i)}(\omega) \propto \omega
\frac{\exp \beta \omega +1}{\exp \beta \omega -1}.
\end{eqnarray}
This relates the quantities ${\cal M}^{(i)}$, 
which are the coefficients in the Langevin equation, with 
the quantities $I_{i}$, which characterizes the fluctuation force field.
This is the generalized fluctuation-dissipation theorem.
In Refs.~\citen{ref:Hu-Paz-Zha} and \citen{ref:Hu-Paz-Zha2},
it is pointed out that
this relation is equivalent to the well-known Einstein formula
in the classical or high-temperature limit
for the case of quantum Brownian motion.

We now discuss the arbitrariness of the definition 
of the fluctuation force field in IFM calculation.
In the case of IFM, an auxiliary field is introduced 
to define the real stochastic effective action.
It is interpreted as the fluctuation force field.
However, in such a definition, 
this force field cannot be uniquely fixed.
For example, the influence action does not change 
even if the term $\delta I^{-1}_{i}$, which has the property 
\begin{eqnarray}
\delta I^{-1}_{i}(x_{1}-x_{2}) = -\delta I^{-1}_{i}(x_{2}-x_{1}), 
\end{eqnarray}
is added to $I^{-1}_{i}$, because of the property
\begin{eqnarray}
\int^t_{t_{0}}d^4 x_{1} d^4 x_{2} \xi_{i}(x_{1})\delta
I^{-1}_{i}(x_{1}-x_{2})\xi_{i}(x_{2}) =0.
\end{eqnarray}

Another point of arbitrariness 
in the definition of the fluctuation force field is that it 
is possible to use only one kind of auxiliary field 
instead of three.\cite{ref:Rish}
In this case, the stochastic influence action is given as
\begin{eqnarray}
\tilde{S}_{IF}[\phi_{<},\phi'_{<};\xi'] = {\rm Re}[S_{IF}[\phi_{<},\phi'_{<}]]
+\int^{t}_{t_{0}}d^4 x
(\phi_{<}(x)-\phi'_{<}(x))\xi'(x),
\end{eqnarray}
and the stochastic weight is 
\begin{eqnarray}
P'[\xi'] &=& N\exp \left\{ \int^{t}_{t_{0}}d^4 x_{1}d^4 x_{2}\xi'(x_{1})I^{-1}(x_{1}-x_{2})\xi'(x_{2}) \right \},
\end{eqnarray}
where
\begin{eqnarray}
I^{-1}(x_{1}-x_{2})
&=&
I^{-1}_{1}(x_{1}-x_{2})
+\bar{\phi}(x_{1})I^{-1}_{2}(x_{1}-x_{2})\bar{\phi}(x_{2}) \nonumber \\
&&+[\bar{\phi}^2(x_{1})+ \frac{1}{12}\phi_{\Delta}^2(x_{1})]
I^{-1}_{3}(x_{1}-x_{2})
[\bar{\phi}^2(x_{2})+ \frac{1}{12}\phi_{\Delta}^2(x_{2})] \nonumber \\
&=& I^{-1}_{1}(x_{1}-x_{2})+\phi_{<}(x_{1})I^{-1}_{2}(x_{1}-x_{2})\phi_{<}(x_{2}) \nonumber \\
&&+\phi^2_{<}(x_{1})
I^{-1}_{3}(x_{1}-x_{2})\phi^2_{<}(x_{2}).
\end{eqnarray}
In the last line, we have taken the limit $\phi_{\Delta}(x)\rightarrow 0$.
The derived semiclassical Langevin equation is
\begin{eqnarray}
&&\Box \phi_{<}({\bf x},t) + m^{2}\phi_{<}({\bf x},t) + \frac{\lambda}{6}\phi_{<}^3({\bf x},t)
 + \left.
      \sum_{i=a,b,c,d,e}(-1)\frac{\delta{\rm Re}[S^{(i)}_{IF}[\phi_{<},\phi'_{<}]]}{\delta \phi_{\Delta}} \right|_{\phi_{\Delta=0}} \nonumber \\
&&\hspace*{4cm} =\xi'({\bf x},t).
\end{eqnarray}
Now, we have only one kind of fluctuation force field.
Therefore, the fluctuation force field introduced in Eq.~(\ref{eqn:f-v-phi4}) 
is only one of several possibilities.
There is still another point of arbitrariness, as 
is discussed in the next section.
It plays an important role when we compare 
the quantum Langevin equation derived using POM with 
the semiclassical Langevin equation derived using IFM.

\section{Quantum Langevin equation in POM 
and semiclassical Langevin equation in IFM}

To this point, we have applied POM and IFM to $\phi^4$ theory and have derived
two Langevin equations, (\ref{eqn:p-phi4}) and (\ref{eqn:f-v-phi4}).
The equation derived with POM is the quantum Langevin equation, 
while that derived with IFM is the semiclassical Langevin equation.
It is useful to compare 
the POM quantum Langevin equation 
with the IFM semiclassical Langevin equation 
and study the correspondence between them.
For the remainder of this section, we ignore the difference between
operators and c-numbers for the soft modes.
More precisely, we make the identification $[\hat{A},\hat{B}]_{+} = 2AB$, 
where $A$ and $B$ are appropriate c-numbers corresponding to $\hat{A}$ 
and $\hat{B}$, respectively.

First, we compare the l.h.s. of the two equations.
It is easily seen 
that there are two differences between the two equations. 
The first difference is that in POM, there are additional contributions 
that correspond to the diagrams shown in Figs.1 (g) and (h), 
which contain the propagator of the soft modes.
Such terms are not included in IFM equation, because 
the quantum effects from the soft modes are omitted there.

The second difference is that 
between the cutoff delta function and the Dirac delta function.
The origin of this difference is
the definition of the functional derivative, which
is used to derive the Langevin equation from the stochastic effective action
in IFM.
Greiner et al. \cite{ref:Grei-Mul} 
used the usual definition of the functional derivative,
\begin{eqnarray}
\frac{\delta\phi_{<}(x')}{\delta\phi_{<}(x)} &=&
\lim_{\epsilon \rightarrow 0}
\frac{[\phi_{<}(x')+\epsilon\delta^{(4)}(x'- x)]-\phi_{<}(x')}{\epsilon}
\nonumber \\
&=& \delta^{(4)}(x'- x).
\label{eqn:fd-o}
\end{eqnarray}
In this definition,
the function with small variation $\phi_{<}(x')+\epsilon\delta^{(4)}(x'- x)$
includes the hard modes.
This is unsatisfactory because, in our method of calculation, 
all the fields that contain the hard components
should be integrated or projected out.
This point can be confirmed by comparing the Fourier components 
of the two Langevin equations.
We consider, for example, the third terms on the l.h.s. of the two equations.
In IFM, the ${\bf k}$ component of the third term is
\begin{eqnarray}
\lefteqn{\frac{\lambda}{6}\int d^3 {\bf x}\phi^3_{<}({\bf x}) e^{-i{\bf kx}}} \nonumber \\
&&= \frac{\lambda}{6(2\pi)^6}\int^{\Lambda_{I}}_{0}d^3 {\bf k}_{1}d^3 {\bf k}_{2}
    \theta(\Lambda_{I}-|{\bf k}-{\bf k}_{1}-{\bf k}_{2}|)
    \phi_{<}({\bf k}_{1})\phi_{<}({\bf k}_{2})\phi_{<}({\bf k}-{\bf k}_{1}-{\bf k}_{2}), \nonumber \\
\label{eqn:3rd-k-gm}
\end{eqnarray}
and in POM, it is
\begin{eqnarray}
\lefteqn{\frac{\lambda}{6}\int d^3 {\bf x}\int d^3 {\bf y} \delta^{(3)}_{\Lambda_{I}}({\bf x}-{\bf y})\phi^3_{<}({\bf y}) e^{-i{\bf kx}} } \nonumber \\
&&=  \frac{\lambda}{6(2\pi)^6}\int^{\Lambda_{I}}_{0}d^3 {\bf k}_{1}d^3 {\bf k}_{2}
    \theta(\Lambda_{I}-|{\bf k}|)\theta(\Lambda_{I}-|{\bf k}-{\bf
    k}_{1}-{\bf k}_{2}|) \nonumber \\
&&\times    \phi_{<}({\bf k}_{1})\phi_{<}({\bf k}_{2})\phi_{<}({\bf k}-{\bf k}_{1}-{\bf k}_{2}). \nonumber \\
\label{eqn:3rd-k-pom}
\end{eqnarray}
Here, the time argument of $\phi_{<}$ has been omitted.
The only difference between 
Eq.~(\ref{eqn:3rd-k-gm}) and Eq.~(\ref{eqn:3rd-k-pom})
is the existence of the step function $\theta(\Lambda_{I}-|{\bf k}|)$
in Eq.~(\ref{eqn:3rd-k-pom}).
The absolute values of the momenta ${\bf k}_{1},{\bf k}_{2}$
and ${\bf k}-{\bf k}_{1}-{\bf k}_{2}$ are always smaller
than $\Lambda_{I}$ in both cases.
However, the situation is different in the case of the momentum ${\bf k}$.
In POM, only components smaller than $\Lambda_{I}$ are included, 
even for momentum ${\bf k}$.
Contrastingly, in IFM, 
all components satisfying the condition 
$0\le |{\bf k}| \le 3\Lambda_{I}$ are included, because the 
step function $\theta(\Lambda_{I}-|{\bf k}|)$ does not appear.
This means that the separation of soft modes and hard modes is
incomplete in the IFM calculation.

To solve this problem, 
we use the following definition
of the functional derivative:
\begin{eqnarray}
\frac{\delta\phi_{<}(x')}{\delta\phi_{<}(x)} =
\delta(x_{0}'-x_{0})\delta^{(3)}_{\Lambda_{I}}({\bf x}'-{\bf x}),
\label{eqn:fd-n}
\end{eqnarray}
where $\delta^{(3)}_{\Lambda_{I}}({\bf x}'-{\bf x})$
is the cutoff delta function defined in Eq.~(\ref{eqn:l-delta}).
If this new definition is adopted in IFM, 
the l.h.s. of the two equations coincide, except for 
the terms corresponding to (g) and (h) in Fig. 1.

Now, we compare the r.h.s. of the two equations.
If we use the definition (\ref{eqn:fd-n}) instead of 
Eq.~(\ref{eqn:g-m-henbun}), 
we obtain Eq.~(\ref{eqn:f-v-phi4}) with the r.h.s. replaced by
\begin{eqnarray}
\int d^3 {\bf y}\{ \xi_{1}({\bf y},t) + \phi_{<}({\bf y},t)\xi_{2}({\bf y},t) + \phi_{<}^2({\bf y},t) \xi_{3}({\bf y},t) \}
\delta^{(3)}_{\Lambda_{I}}({\bf x}-{\bf y}).
\end{eqnarray}
Comparing this with the r.h.s. of Eq.~(\ref{eqn:p-phi4}), 
we see that there are still the following differences between them.
First, the time dependences 
of the soft modes that appear on the r.h.s.~of the two equations 
are different.
In IFM, this time dependence is determined by solving the equation, 
while in POM, they evolve as free fields.
This is a considerable difference.
In POM, the r.h.s.~of the quantum Langevin equation vanishes 
when we calculate the expectation value 
with the initial density matrix, because the soft modes cannot include
the $\hat{f}_{i}$ dependence.
By contrast, in IFM, the soft modes included 
in the r.h.s.~of the semiclassical Langevin equation 
have a $\xi_{i}$ dependence, 
and therefore the r.h.s.~does not necessarily become zero when we 
calculate the expectation value with stochastic weights; that is, 
\begin{eqnarray}
&&\langle \hat{f}_{i}({\bf x},t)\hat{\phi}^{n}_{0<}({\bf x},t) \rangle = 0, \\
&&\langle \xi_{i}({\bf x},t)\phi^{n}_{<}({\bf x},t) \rangle_{P}
= \langle \xi_{i}({\bf x},t)\phi^{n}_{<}({\bf x},t;\xi_{i})
\rangle_{P} 
\neq 0.
\end{eqnarray}
Second, the correlations of the fluctuation force fields are different.
Consider, for example, the second order correlations 
$\langle \xi_{i}({\bf x}_{1},t_{1})\xi_{j}({\bf x}_{2},t_{2}) \rangle_{P}$
and $\langle \hat{f}_{i}({\bf x}_{1},t_{1})
\hat{f}_{j}({\bf x}_{2},t_{2}) \rangle$.
The former is real and invariant under
the exchange of the arguments
${\bf x}_{1}\leftrightarrow {\bf x}_{2},t_{1}\leftrightarrow t_{2}$, 
as seen from
Eqs.~(\ref{eqn:xi11}), (\ref{eqn:xi22}) and (\ref{eqn:xi33}).
The latter is complex and is not invariant under the same exchange, 
as seen from 
Eqs.~(\ref{eqn:flu-1}), (\ref{eqn:flu-2}) and (\ref{eqn:flu-3}).
Therefore, the corresponding correlations in POM 
are considered to be the symmetrized correlations
given by the real parts of the second-order correlations defined
in Eqs.~(\ref{eqn:f11}),~(\ref{eqn:f22}) and (\ref{eqn:f33}).
Then, we notice that Eqs.~(\ref{eqn:xi22}) and (\ref{eqn:xi33})
are identical to Eqs.~(\ref{eqn:f22}) and (\ref{eqn:f33}), respectively:
\begin{eqnarray}
\langle [\hat{f}_{2}({\bf x},t),\hat{f}_{2}({\bf x}',t')]_{+} \rangle /2
&=& \langle \xi_{2}({\bf x},t)\xi_{2}({\bf x}',t') \rangle_{P}, \\
\langle [\hat{f}_{3}({\bf x},t),\hat{f}_{3}({\bf x}',t')]_{+} \rangle /2
&=& \langle \xi_{3}({\bf x},t)\xi_{3}({\bf x}',t') \rangle_{P}.
\end{eqnarray}
Contrastingly, Eq.~(\ref{eqn:xi11}) differs from Eq.~(\ref{eqn:f11}) 
because of the existence of the additional term $\delta I_{1}$:
\begin{eqnarray}
\langle [\hat{f}_{1}({\bf x},t),\hat{f}_{1}({\bf x}',t')]_{+} \rangle/2
= \langle \xi_{1}({\bf x},t)\xi_{1}({\bf x}',t') \rangle_{P}
-\frac{1}{2}\delta I_{1}({\bf x}-{\bf x}',t-t'),
\end{eqnarray}
where
\begin{eqnarray}
\delta I_{1}({\bf x}-{\bf x}',t-t')
&=& -i\frac{\lambda^2}{4}G_{\Lambda_{I}}(0,0)^2
[G_{\Lambda_{I}}({\bf x}-{\bf x}',t-t') +G_{\Lambda_{I}}({\bf x}'-{\bf x},t'-t) ] \nonumber \\
&=& \frac{\lambda^2}{2}G^{\Lambda_{I}}_{++}(0)^2
{\rm Im}
G^{\Lambda_{I}}_{++}(x-x').
\end{eqnarray}
However, this difference can be removed by modifying $\xi_{1}$.
First, note that $\delta I_{1}$ has the following property 
\begin{eqnarray}
\int d^4 x_{1}d^4 x_{2} \phi_{<}(x_{1})
\delta I^{-1}_{1}({\bf x}_{1}-{\bf x}_{2},t_{1}-t_{2}) \phi_{<}(x_{2}) = 0.
\end{eqnarray}
This property can be proved using the fact that 
$\phi_{<}(x)$ contains only soft modes, 
while $\delta I^{-1}_{1}$ contains only hard modes.
With this fact, 
the imaginary part of the influence action (\ref{eqn:IA-IM}) 
can be rewritten as 
\begin{eqnarray}
\lefteqn{{\rm Im}S_{IF}[\phi_{<},\phi'_{<}]} && \nonumber \\
&\rightarrow& \lambda^2 \int^{t}_{t_{0}}d^4 x_{1} d^4 x_{2}
\left\{
\phi_{\Delta}(x_{1})
\left( -\frac{1}{12}{\rm Im}[G^{\Lambda_{I}}_{++}(x_{1}-x_{2})^3] -\frac{1}{8}G^{\Lambda_{I}}_{++}(0)^2 {\rm Im}G^{\Lambda_{I}}_{++}(x_{1}-x_{2}) \right)
\phi_{\Delta}(x_{2}) \right. \nonumber \\
&& \left. -\frac{1}{4}\phi_{\Delta}(x_{1})\bar{\phi}(x_{1})
{\rm Re}
[G^{\Lambda_{I}}_{++}(x_{1}-x_{2})^2]
\phi_{\Delta}(x_{2})\bar{\phi}(x_{2}) \right.
\nonumber \\
&& \left. +\frac{1}{8}\phi_{\Delta}(x_{1})[\bar{\phi}^2(x_{1}) +\frac{1}{12}\phi^2_{\Delta}(x_{1})]
{\rm Im}
[G^{\Lambda_{I}}_{++}(x_{1}-x_{2})]
\phi_{\Delta}(x_{2})[\bar{\phi}^2(x_{2}) + \frac{1}{12}\phi_{\Delta}^2(x_{2})]
\right\}.
\end{eqnarray}
This imaginary part can be reproduced by replacing $\xi_{1}$ in 
the stochastic effective action (\ref{eqn:sto-eac}) with 
the auxiliary field $\xi'_{1}$ with the stochastic weight
\begin{eqnarray}
P_{1}[\xi'_{1}] &=& N'_{1}\exp
\left\{ \int^{t}_{t_{0}}d^4 x_{1}d^4
x_{2}\xi'_{1}(x_{1})(I^{-1}_{1}(x_{1}-x_{2})+\delta I^{-1}_{1}(x_{1}-x_{2}))\xi'_{1}(x_{2})
\right \}. \nonumber \\
\end{eqnarray}
The second-order correlation of $\xi'_{1}$ then becomes 
\begin{eqnarray}
\langle \xi'_{1}({\bf x},t) \xi'_{1}({\bf x}',t') \rangle_{P}
&=& -\frac{1}{2}(I_{1}(x_{1}-x_{2})+\delta I_{1}(x_{1}-x_{2}))
\nonumber \\
&=& \langle [\hat{f}_{1}({\bf x},t),\hat{f}_{1}({\bf x}',t')]_{+} \rangle/2 .
\end{eqnarray}
In short, we have shown that the second-order correlations in IFM 
coincide with those in POM by utilizing the arbitrariness of the
definition of the fluctuation force field in IFM.
However, it should be noted that 
in POM, there is also the nonzero correlation 
\begin{eqnarray}
\langle [\hat{f}_{1}({\bf x}_{1},t_{1}),\hat{f}_{3}({\bf x}_{2},t_{2})]_{+} \rangle/2
&=&
-\frac{\lambda^2}{8}
G_{\Lambda_{I}}(0,0)
[G_{\Lambda_{I}}({\bf x}_{1}-{\bf x}_{2},t_{1}-t_{2}) +G_{\Lambda_{I}}({\bf x}_{2}-{\bf x}_{1},t_{2}-t_{1}) ]. \nonumber \\
\label{eqn:f13}
\end{eqnarray}
The corresponding correlation is implicitly assumed to vanish in IFM.

Now, we briefly discuss the fluctuation-dissipation theorem.
In IFM, it is pointed out that the generalized f-d theorem is satisfied.
This theorem asserts that there is a relation between the quantities 
${\cal M}^{(i)}$, which are defined by 
Eqs.~(\ref{eqn:m-11}), (\ref{eqn:m-22}) and (\ref{eqn:m-33}), 
and the quantities $I_{i}$ that determines 
the second-order correlation of the fluctuation force field.
In order to see whether such a relation is satisfied in POM,
it is necessary to identify the corresponding quantities 
${\cal M}^{(i)}$ and $I_{i}$ in POM.
From the fact that the l.h.s.~of the two equations have the same form,
it is seem that the ${\cal M}^{(i)}$ in IFM 
are the same as those in POM.
On the other hand, the correspondence between the r.h.s.
of the two equations is not clear.
We assume that $I_{1}$, $I_{2}$ and $I_{3}$ in POM correspond to 
the l.h.s. (or the r.h.s.) of 
Eqs.~(\ref{eqn:f11}), (\ref{eqn:f22}) and (\ref{eqn:f33}), respectively.
Equations (\ref{eqn:f22}) and (\ref{eqn:f33}) are proportional
to $I_{2}$ and $I_{3}$ given in IFM.
Therefore, the generalized f-d theorem is satisfied for them.
However, the r.h.s. of Eq. (\ref{eqn:f11}) is not proportional to $I_{1}$, 
and therefore, the generalized f-d theorem is not satisfied
for the $\hat{f}_{1}$ part.

In POM, there is a relation (the 2nd f-d theorem) 
between the fluctuation force fields and 
the coefficients in the quantum Langevin equation.
There is no such relation in IFM, because the imaginary part
of the second-order correlations is absent.
In other words, 
the fluctuation force field in POM possesses more information than that in IFM.

\section{Derivation of a semiclassical Langevin equation in POM}
\label{chap:com-cla-pom}

In this section, we derive a semiclassical Langevin equation 
from the quantum Langevin equation (\ref{eqn:p-phi4}).

First, we must remove quantum effects that come from the system.
For this purpose, we replace the operators of the soft modes with c-numbers.
There are two possible methods for this replacement.
In the first, 
we make this replacement after taking the expectation value using the 
initial density matrix:
\begin{eqnarray}
\hat{\phi}_{<}({\bf x}_{1},t_{1})\hat{\phi}_{<}({\bf x}_{2},t_{2})\cdots
\hat{\phi}_{<}({\bf x}_{n},t_{n})
&\longrightarrow&
{\rm Tr} [\rho \hat{\phi}_{<}({\bf x}_{1},t_{1})
\hat{\phi}_{<}({\bf x}_{2},t_{2})\cdots\hat{\phi}_{<}({\bf
x}_{n},t_{n})] \nonumber \\
&\sim&
\phi_{<}({\bf x}_{1},t_{1})\phi_{<}({\bf x}_{2},t_{2})\cdots
\phi_{<}({\bf x}_{n},t_{n}),
\end{eqnarray}
for $n=1,2,3,\cdots$.
Here $\rho$ is the initial density matrix of the total system.
In the second, we replace the operators themselves with the c-numbers:
\begin{eqnarray}
\hat{\phi}_{<}({\bf x}_{1},t_{1})\hat{\phi}_{<}({\bf x}_{2},t_{2})\cdots
\hat{\phi}_{<}({\bf x}_{n},t_{n})
\longrightarrow
\phi_{<}({\bf x}_{1},t_{1})\phi_{<}({\bf x}_{2},t_{2})\cdots
\phi_{<}({\bf x}_{n},t_{n}).\nonumber \\
\end{eqnarray}
In the former method of replacement, 
we cannot obtain the fluctuation force field, because
the r.h.s. of the equation becomes zero when we take the expectation value, 
as follows from its definition.
For this reason, we adopt the latter method of replacement.
Note that a similar replacement should be made for 
$\hat{\phi}_{0<}({\bf x},t)$ and $\hat{\phi}_{0<}^2({\bf x},t)$ 
in the fluctuation force field.
Furthermore, in Eq.~(\ref{eqn:p-phi4}), there are two terms 
including the soft mode propagator, 
which correspond to the diagrams shown in Figs.~1 (g) and (h).
In deriving our semiclassical equation, we simply drop these terms.

Next, we have to replace the fluctuation force field with a c-number.
The simplest method to do so is 
to introduce c-number fields $f_{1}$, $f_{2}$ and $f_{3}$ that have 
the following correlations:
\begin{eqnarray}
&& \langle f_{1}({\bf x},t) \rangle_{c} =
 \langle f_{2}({\bf x},t)  \rangle_{c} =
  \langle f_{3}({\bf x},t)  \rangle_{c} = 0. \\
&& \langle f_{1}({\bf x}_{1},t_{1})f_{1}({\bf x}_{2},t_{2}) \rangle_{c}
= -\frac{\lambda^2}{12i}
[G_{\Lambda_{I}}({\bf x}_{1}-{\bf x}_{2},t_{1}-t_{2})^3 +G_{\Lambda_{I}}({\bf x}_{2}-{\bf x}_{1},t_{2}-t_{1})^3 ]
 \nonumber \\
&& \hspace*{2cm}+\frac{i\lambda^2}{8}G_{\Lambda_{I}}(0,0)^2
[G_{\Lambda_{I}}({\bf x}_{1}-{\bf x}_{2},t_{1}-t_{2}) +G_{\Lambda_{I}}({\bf x}_{2}-{\bf x}_{1},t_{2}-t_{1}) ]
, \\
&&\langle f_{2}({\bf x}_{1},t_{1})f_{2}({\bf x}_{2},t_{2}) \rangle_{c}
= -\frac{\lambda^2}{4}
[ G_{\Lambda_{I}}({\bf x}_{1}-{\bf x}_{2},t_{1}-t_{2})^2 + G_{\Lambda_{I}}({\bf x}_{2}-{\bf x}_{1},t_{2}-t_{1})^2 ]
 , \\
&&\langle f_{3}({\bf x}_{1},t_{1})f_{3}({\bf x}_{2},t_{2}) \rangle_{c}
= \frac{\lambda^2}{8i}
[ G_{\Lambda_{I}}({\bf x}_{1}-{\bf x}_{2},t_{1}-t_{2}) + G_{\Lambda_{I}}({\bf x}_{2}-{\bf x}_{1},t_{2}-t_{1}) ]
, \\
&&\langle f_{1}({\bf x}_{1},t_{1})f_{3}({\bf x}_{2},t_{2}) \rangle_{c}
= -\frac{\lambda^2}{8}
G_{\Lambda_{I}}(0,0)
[G_{\Lambda_{I}}({\bf x}_{1}-{\bf x}_{2},t_{1}-t_{2}) +G_{\Lambda_{I}}({\bf x}_{2}-{\bf x}_{1},t_{2}-t_{1}) ].
\nonumber \\
\end{eqnarray}
Here, $\langle~\rangle_{c}$ represents an average evaluated with
respect to suitable stochastic weights.
Other second-order correlations are assumed to be zero.
These values are determined so as to reproduce the real part 
of the second-order correlations, in other word, the symmetrized 
correlations of $\hat{f}_{i}$.
If there are no correlations between different $f_{i}$, 
we obtain Gaussian stochastic weights, as in case of IFM.
Actually, there is a correlation between $f_{1}$ and $f_{3}$ in this case, 
and it is not clear whether or not one can construct a stochastic 
weight that reproduces all the above correlations.

Because of the above stated problem, 
we consider another replacement, which is explicitly 
defined in terms of a definite stochastic weight.
In the IFM calculation, 
the field $\xi_{i}$ is treated as the stochastic variable.
We regard the creation-annihilation operator as the stochastic variable.
The creation-annihilation operators of the hard modes, which develop 
as free fields, have the correlation 
\begin{eqnarray}
\langle a^{\dagger}_{\bf k}(t)a_{\bf k'}(t') \rangle
&=& {\rm Tr}_{E}[\rho_{E}a^{\dagger}_{\bf k}a_{\bf k'}e^{i\omega_{\bf k}t}e^{-i\omega_{\bf k'}t'}] \nonumber \\
&=& n_{\omega_{\bf k}}e^{i\omega_{\bf k}(t-t')}\delta^{(3)}({\bf k-k'})
\nonumber \\
&\equiv& n({\bf k},{\bf k}';t,t')
\end{eqnarray}
We consider the replacement of the creation-annihilation operators 
of the hard modes 
$a_{\bf k}(t)$ and $a^{\dagger}_{\bf k}(t)$ with the c-numbers 
$\beta_{\bf k}(t)$ and $\beta^{*}_{\bf k}(t)$, respectively.
Then, to realize the above correlation for these c-numbers, 
we introduce the stochastic weight $P[\beta,\beta^{*}]$ which has the property 
\begin{eqnarray}
\langle \beta^{*}_{\bf k}(t)\beta_{\bf k'}(t')\rangle_{\beta}
&=& \prod_{{\bf k}\ge \Lambda_{I}}\int D\beta_{\bf k}D\beta^*_{\bf k}
P[\beta,\beta^*]\beta^{*}_{\bf k}(t)\beta_{\bf k'}(t') \nonumber \\
&=&n({\bf k},{\bf k}';t,t'),
\end{eqnarray}
where
\begin{eqnarray}
P[\beta,\beta^*] =
N\exp\left[ -\int^{t}_{t_{0}}ds_{1}ds_{2}\int d^3 {\bf k}_{1}d^3 {\bf k}_{2}
\beta_{{\bf k}_{1}}(s_{1})
n^{-1}({\bf k}_{1},{\bf k}_{2};s_{1},s_{2})
\beta^{*}_{{\bf k}_{2}}(s_{2}) \right]. \nonumber \\
\end{eqnarray}
In this replacement, 
the field operator $\hat{\phi}_{0>}({\bf x},t)$ is replaced with 
the c-number field $\phi_{>}(\beta,\beta^*,{\bf x},t)$, given by
\begin{eqnarray}
\phi_{>}(\beta,\beta^*,{\bf x},t)
&=& \int^{\Lambda}_{\Lambda_{I}}d^3{\bf k}
\frac{1}{\sqrt{2(2\pi)^3 \omega_{\bf k}}}
[\beta_{\bf k}(t)e^{i{\bf kx}} + \beta^{*}_{\bf k}(t)e^{-i{\bf kx}} ].
\end{eqnarray}
Using this function $\phi_{>}$, 
we replace $\hat{f}_{i}$ with c-numbers as follows: 
\begin{eqnarray}
\hat{f}_{1}({\bf x},t) &\longrightarrow &
f_{1}(\beta,\beta^{*},{\bf x},t)=
\frac{\lambda}{6} \phi_{>}(\beta,\beta^*,{\bf x},t)^3, \\
\hat{f}_{2}({\bf x},t) &\longrightarrow &
f_{2}(\beta,\beta^{*},{\bf x},t)=
\frac{\lambda}{2}
(\phi_{>}(\beta,\beta^*,{\bf x},t)^2 -\langle \phi_{>}(\beta,\beta^{*},{\bf x},t)^2 \rangle_{\beta} )
, \\
\hat{f}_{3}({\bf x},t) &\longrightarrow &
f_{3}(\beta,\beta^{*},{\bf x},t) =
\frac{\lambda}{2}\phi_{>}(\beta,\beta^*,{\bf x},t),
\end{eqnarray}

Summarizing the above result, the semiclassical Langevin equation 
in POM is given by
\begin{eqnarray}
\Box \phi_{<}({\bf x},t) + m^2 \phi_{<}({\bf x},t)
+ \sum_{i=a,\cdots,f}\int d^3 {\bf y}
 \delta^{(3)}_{\Lambda_{I}}({\bf x}-{\bf y})
G_{i}[\phi_{<}({\bf y},t)] &&\nonumber \\
&&\hspace*{-10cm}= \int d^3 {\bf y}
\delta^{(3)}_{\Lambda_{I}}({\bf x}-{\bf y})
\{ f_{1}(\beta,\beta^{*},{\bf y},t) + \phi_{0<}({\bf y},t)
 f_{2}(\beta,\beta^{*}{\bf y},t) + \phi_{0<}^2({\bf y},t)
 f_{3}(\beta,\beta^{*}{\bf y},t) \}. \nonumber \\
 \label{eqn:phi4pom-cla} 
\end{eqnarray}
For comparison, the semiclassical equation in IFM is given by
\begin{eqnarray} 
\Box \phi_{<}({\bf x},t) + m^2 \phi_{<}({\bf x},t) 
+ \sum_{i=a,\cdots,f}\int d^3 {\bf y} \delta^{(3)}({\bf x}-{\bf y}) 
G_{i}[\phi_{<}({\bf y},t)] &&\nonumber \\ 
&&\hspace*{-10cm}= \int d^3 {\bf y}
\delta^{(3)}({\bf x}-{\bf y}) 
\{ \xi_{1}({\bf y},t) + \phi_{<}({\bf y},t) \xi_{2}
({\bf y},t) + \phi_{<}^2({\bf y},t)
\xi_{3}({\bf y},t) \}, \nonumber \\
\label{eqn:phi4ifm-cla}
\end{eqnarray}
where
\begin{eqnarray}
G_{a}[\phi_{<}({\bf y},t)]
&=&
\frac{\lambda}{6}
\phi_{<}^3({\bf y},t), \\
G_{b}[\phi_{<}({\bf y},t)]
&=&
-\frac{i\lambda}{2}G_{\Lambda_{I}}(0,0) \phi_{<}({\bf y},t),\\
G_{c}[\phi_{<}({\bf y},t)]
&=&
-\frac{\lambda^2}{4}\int^{t}_{0}ds \int d^3 {\bf y}_{1}
[G_{\Lambda_{I}}({\bf y}_{1}-{\bf y},s-t)^2 - G_{\Lambda_{I}}({\bf y}-{\bf y}_{1},t-s)^2] G_{\Lambda_{I}}(0,0) \phi_{<}({\bf y},t), \nonumber \\
\\
G_{d}[\phi_{<}({\bf y},t)]
&=&
 -\frac{\lambda^2}{6}\int^{t}_{0}ds \int d^3 {\bf y}_{1}
[G_{\Lambda_{I}}({\bf y}_{1}-{\bf y},-s)^3 -G_{\Lambda_{I}}({\bf y}-{\bf y}_{1},s)^3]
\varphi_{<}({\bf y}_{1},-s,t), \\
G_{e}[\phi_{<}({\bf y},t)]
&=&
-\frac{i\lambda^2}{4}\int^{t}_{0}ds \int d^3 {\bf y}_{1}
[G_{\Lambda_{I}}({\bf y}_{1}-{\bf y},-s)^2 -G_{\Lambda_{I}}({\bf y}-{\bf y}_{1},s)^2]
\varphi_{<}^{2}({\bf y}_{1},-s,t)\phi_{<}({\bf y},t), \nonumber \\
\\
G_{f}[\phi_{<}({\bf y},t)]
&=&
\frac{\lambda^2}{12}\int^{t}_{0}ds \int d^3 {\bf y}_{1}
[G_{\Lambda_{I}}({\bf y}_{1}-{\bf y},-s) -G_{\Lambda_{I}}({\bf y}-{\bf y}_{1},s)]
\varphi_{<}^{3}({\bf y}_{1},-s,t)\phi_{<}^2 ({\bf y},t). \nonumber \\
\end{eqnarray}
It is clear from the discussion in the previous section that 
the l.h.s.~of the two equations are identical except for 
the difference between the Dirac delta function 
and the cutoff delta function.
For the fluctuation force fields, 
which are the r.h.s. of the equations,  
there is a difference between POM and IFM at the level of the 
semiclassical Langevin equation:
In IFM, 
$\xi_{1}$, $\xi_{2}$ and $\xi_{3}$ are treated as stochastic 
variables, while in POM, 
there are only two stochastic variables $\beta$ and $\beta^*$.
As a result, there are some differences in the correlations.
To show this, 
we calculate the correlations of $f_{i}(\beta,\beta^{*},{\bf x},t)$.

The first order correlations are
\begin{eqnarray}
\langle f_{1}(\beta,\beta^{*},{\bf x},t)  \rangle_{\beta}
= \langle f_{2}(\beta,\beta^{*},{\bf x},t)  \rangle_{\beta}
= \langle f_{3}(\beta,\beta^{*},{\bf x},t)  \rangle_{\beta}
=0.
\end{eqnarray}
Thus all the first-order correlations disappear.
This is also the case in IFM.
The second order correlations are 
\begin{eqnarray}
\lefteqn{ \langle f_{1}(\beta,\beta^{*},{\bf x}_{1},t_{1})f_{1}(\beta,\beta^{*},{\bf x}_{2},t_{2}) \rangle_{\beta} } && \nonumber \\
&=& \frac{4\lambda^2}{3}
\int^{\Lambda}_{\Lambda_{I}}d^3 {\bf l}_{1} d^3{\bf l}_{2} d^3{\bf l}_{3}
\frac{1}{8(2\pi)^9 \omega_{{\bf l}_{1}}\omega_{{\bf l}_{2}} \omega_{{\bf l}_{3}}}n_{\omega_{{\bf l}_{1}}}n_{\omega_{{\bf l}_{2}}}
n_{\omega_{{\bf l}_{3}}} \nonumber \\
&&\times \cos \omega_{{\bf l}_{1}}(t_{1}-t_{2})
\cos \omega_{{\bf l}_{2}}(t_{1}-t_{2})
\cos \omega_{{\bf l}_{3}}(t_{1}-t_{2}) 
e^{i({\bf l}_{1}+{\bf l}_{2}+{\bf l}_{3})({\bf x}_{1}-{\bf x}_{2})} \nonumber \\
&& + 2\lambda^2
\int^{\Lambda}_{\Lambda_{I}}d^3 {\bf l}_{1} d^3{\bf l}_{2} d^3{\bf l}_{3}
\frac{1}{8(2\pi)^9 \omega_{{\bf l}_{1}}\omega_{{\bf l}_{2}}\omega_{{\bf l}_{3}}}
n_{\omega_{{\bf l}_{1}}}n_{\omega_{{\bf l}_{2}}}n_{\omega_{{\bf l}_{3}}}
\cos \omega_{{\bf l}_{3}}(t_{1}-t_{2}) 
e^{i{\bf l}_{3}({\bf x}_{1}-{\bf x}_{2})}, \nonumber \\
\label{eqn:2nd-flu-cla-11}\\
\lefteqn{ \langle f_{2}(\beta,\beta^{*},{\bf x}_{1},t_{1}) f_{2}(\beta,\beta^{*},{\bf x}_{2},t_{2}) \rangle_{\beta} } && \nonumber \\
&=&
2\lambda^2 \int^{\Lambda}_{\Lambda_{I}}d^3 {\bf l}_{1} d^3{\bf l}_{2}
\frac{1}{4(2\pi)^6 \omega_{{\bf l}_{1}}\omega_{{\bf l}_{2}}}
n_{\omega_{{\bf l}_{1}}}n_{\omega_{{\bf l}_{2}}} \nonumber \\
&&\times \cos \omega_{{\bf
l}_{1}}(t_{1}-t_{2})\cos \omega_{{\bf l}_{2}}(t_{1}-t_{2})e^{i({\bf l}_{1}+{\bf l}_{2})({\bf x}_{1}-{\bf x}_{2})},\label{eqn:2nd-flu-cla-22} \\ \lefteqn{ \langle f_{3}(\beta,\beta^*,{\bf x}_{1},t_{1})f_{3}(\beta,\beta^*,{\bf x}_{2},t_{2}) \rangle_{\beta} } && \nonumber \\
&=& \prod_{{\bf k} \ge \Lambda_{I}} \int \frac{d^2 \beta_{\bf k}}{\pi}
P(\beta_{\bf k})   f_{3}(\beta,\beta^*,{\bf x}_{1},t_{1})
f_{3}(\beta,\beta^*,{\bf x}_{2},t_{2}) \nonumber \\
&=& \frac{\lambda^2}{4}
\int^{\Lambda}_{\Lambda_{I}}d^3 {\bf k}\frac{1}{2(2\pi)^3 \omega_{\bf k}}
2n_{\omega_{\bf k}}\cos \omega_{\bf k}(t_{1}-t_{2})
e^{i{\bf k}({\bf x}_{1}-{\bf x}_{2})},
\label{eqn:2nd-flu-cla-33}\\
\lefteqn{ \langle f_{1}(\beta,\beta^*,{\bf x}_{1},t_{1})f_{3} (\beta,\beta^*,{\bf x}_{2},t_{2}) \rangle_{\beta} } && \nonumber \\
&=& \lambda^2\int^{\Lambda}_{\Lambda_{I}} d^3{\bf l}_{1}d^3 {\bf l}_{2}
\frac{1}{4(2\pi)^6 \omega_{{\bf l}_{1}}\omega_{{\bf l}_{2}} }
n_{\omega_{{\bf l}_{1}}}n_{\omega_{{\bf l}_{2}}}
\cos \omega_{{\bf l}_{1}}(t_{1}-t_{2}) \cos \omega_{{\bf l}_{2}}(t_{1}-t_{2})
e^{i{\bf i}_{2}({\bf x}_{1}-{\bf x}_{2})}. \label{eqn:2nd-flu-cla-13}\nonumber \\
\end{eqnarray}
These correlations are identical with 
the symmetrized second-order correlations if 
$1 + 2 n_{\omega_{\bf p}}$ is replaced with $2 n_{\omega_{\bf p}}$ in 
Eqs.~(\ref{eqn:f11}), (\ref{eqn:f22}), (\ref{eqn:f33}) 
and (\ref{eqn:f13}).
This can be understood as follows. 
The creation-annihilation operators $a_{\bf k}$ and $a^{\dagger}_{\bf k}$ 
depends on their order of the operation:
\begin{eqnarray}
\langle a^{\dagger}_{\bf k}a_{\bf k'} \rangle
\neq \langle a_{\bf k'}a^{\dagger}_{\bf k} \rangle.
\end{eqnarray}
However, in the replacement under consideration, 
the corresponding situation does not exist, as 
\begin{eqnarray}
\langle \beta^{*}_{\bf k}\beta_{\bf k'} \rangle_{\beta}
= \langle \beta_{\bf k'}\beta^{*}_{\bf k} \rangle_{\beta}.
\end{eqnarray}
In other words, in the present replacement,  
all the contributions from the vacuum are ignored.

In this section, we have derived a semiclassical Langevin equation 
from a quantum Langevin equation in POM 
ignoring the noncommutativity of the soft mode operators and 
diagrams with soft mode propagation.
The problem we encounter is that the procedure of replacing the fluctuation 
force fields with c-number stochastic variables is not unique.
With the procedure we employed here, the correct correlations between 
the fluctuation force fields with different indices, 
like Eq.~(\ref{eqn:2nd-flu-cla-13}), are realized 
but the correct contribution from the vacuum is not realized.

\section{Conclusions and discussion}

In this paper, we applied POM to $\phi^4$ theory and derived 
correct quantum and semiclassical Langevin equations 
that describe the time evolution of the soft modes.
We have examined in detail 
the differences and similarities of the results 
obtained using POM and IFM.

In the case of POM, 
the quantum Langevin equation is derived 
systematically from a perturbative expansion 
of the Heisenberg equation of motion.
The fluctuation-dissipation theorem of second kind is applied to 
determine the order of the expansion.
In this formalism, we can derive an equation of motion 
with no time-convolution integral term without 
imposing an approximation, like the linear harmonic approximation.
To derive the semiclassical Langevin equation from this equation of motion, 
the quantum effect arising from the soft modes must be removed 
by replacing the operators of the system, i.e. the soft modes, with 
c-numbers, and we ignored the contribution 
that includes the soft mode propagator.
Furthermore, 
we replaced the fluctuation force field with a c-number 
by regarding the creation-annihilation operators of the hard modes 
as stochastic variables.

In IFM, the semiclassical Langevin equation is derived by applying 
a variational principle to the stochastic effective action.
This equation has time-convolution integral terms, and
the linear harmonic approximation is employed to remove them, while 
maintaining the dissipation effect.
To derive the stochastic effective action, the imaginary part 
of the influence action was removed by introducing auxiliary fields.
These fields are interpreted as the fluctuation force fields.

We compared the semiclassical Langevin equation obtained using IFM with 
the quantum Langevin equation obtained using POM.
If we ignore the fluctuation force field, 
the difference between the two equations comes 
only from the difference between the Dirac delta function 
and the cutoff delta function.
This difference results from the definition of the functional derivative 
that is used to derive 
the semiclassical Langevin equation from the stochastic 
effective action in IFM.
In POM, the momentum component that is larger than the cutoff 
$\Lambda_{I}$ is always projected out, due to this cutoff delta function.
In IFM, by contrast, the separation of the momentum is incomplete.
Note that this agreement of the respective of the l.h.s. is also due to 
the linear harmonic approximation employed in IFM.
There is no such agreement 
if the quasi-instantaneous approximation is used.
In this sense, it may be the case that POM provides the basis for 
the validity of the linear harmonic approximation.

With regard to the fluctuation force field, 
the relation between the two approaches is more complicated.
The soft modes that are coupled to the fluctuation force field in POM 
develop as free fields, and therefore the fluctuation force field term 
always disappears if the expectation value is calculated using the 
initial density matrix.
Contractingly, in IFM, 
the time dependence of the soft modes is 
to be determined by solving the equation.
Therefore, the expectation values of the fluctuation force 
terms in IFM do not vanish in general.

Moreover, in IFM, there is an arbitrariness 
in the auxiliary fields.
It is noteworthy that 
the correlations $\langle \xi_{1}({\bf x},t) \xi_{1}({\bf x}',t') \rangle_{P}$ 
and $\langle [\hat{f}_{1}({\bf x},t),\hat{f}_{1}({\bf x}',t')]_{+} \rangle/2$ 
can be made identical by utilizing this arbitrariness.
In short, we can conclude that in POM, 
we can determine the fluctuation force field uniquely, and 
the result suggests that the fluctuation force field in IFM 
may have to be modified.
In IFM, it is necessary to introduce another principle 
to determine the fluctuation force field uniquely.
Furthermore, it is implicitly assumed that there is no correlation 
between different fluctuation force fields in IFM.
Our POM study indicates that 
such an assumption is not necessarily valid.

In short, the derivation of the Langevin equation in POM
seems to be more systematic and less ambiguous than that in IFM.
Furthermore, the fact that the differences between the two approaches 
can be attributed to ambiguities in IFM seems to indicate the 
superiority of POM.
However, such a claim cannot yet be made conclusively, and 
it is a future problem to study the deeper physical meaning of 
the differences between the two approaches.

Greiner et al. have asserted that the system should thermalize with the 
approximations used in the present paper.
However, there is no common understanding about what condition a
nonlinear Langevin equation should satisfy to ensure the 
evolution of the system toward thermal equilibrium.
The well-known fluctuation-dissipation theorem 
may ensure thermal equilibrium for the linear case, 
but this is an open question for the nonlinear case.\cite{ref:S-H-flu}
Therefore, it is not clear at present under what condition the 
equation of motion obtained with either POM or IFM guarantees the realization 
of thermal equilibrium.
In the present paper, we have compared the two equations 
obtained from POM and IFM with respect to the assumptions 
and arbitrariness contained in each formalism.
On the basis of this comparison, we conclude that POM is better than IFM.

Finally, we derived the semiclassical Langevin equation in POM.
A crucial problem here is to determine 
how to replace the fluctuation force field, 
which consists of an environment operator, with a c-number.
In this paper, we replace the creation-annihilation operators 
of the environment with corresponding to the stochastic variables.
In this replacement, all the vacuum contributions are dropped.
It is a future problem to further study the justification of this replacement.

\appendix

\section{Fluctuation-dissipation theorem of second kind}\label{app:2ndfd}

In conventional statistical mechanics,
the following two relations are known as 
the fluctuation-dissipation theorem.
The first is that the transport coefficient is represented 
in terms of the  fluctuations
of macroscopic variables.
This is called 
the fluctuation-dissipation theorem of first kind (1st f-d theorem).
The second is 
the fluctuation-dissipation theorem of second kind (2nd f-d theorem),
according to which the transport coefficients are 
represented by the fluctuation force.
In POM, the 2nd f-d theorem is applied to determine the order 
of the expansion of the fluctuation force.

The 2nd f-d theorem is usually understood in terms of 
the Mori-Langevin equation,\cite{ref:Mori}
\begin{eqnarray}
\frac{d}{dt}A(t) = i\omega A(t) -\int^{t}_{0}ds \varphi(t-s)A(s) + f(t),
 \label{eqn:ml}
\end{eqnarray}
where $A$ is a macroscopic variable, $\varphi(t)$ the memory function
and $f(t)$ the fluctuation force.
The memory function is given by
\begin{eqnarray}
\varphi(t) = (f(t),f^{\dagger}(0)),  \label{eqn:Mo-Co}
\end{eqnarray}
where $(~~)$ represents an inner product satisfying the following properties:
\begin{eqnarray}
(F,G)^{*} &=& (G,F),  \\
(\sum_{i}c_{i}F_{i},G) &=& \sum_{i}c_{i}(F_{i},G)
\end{eqnarray}
It can be seen in Eq.~(\ref{eqn:Mo-Co}) 
that the memory function is represented by
the time correlation of the fluctuation forces.
This is the 2nd f-d theorem.
From the above discussion, we can see that 
if the memory function is of second order in the coupling constant, 
the fluctuation force must be of 
first order for the 2nd f-d theorem to be satisfied.
The Mori equation is derived exactly from the Heisenberg equation
of motion, and therefore, the 2nd f-d theorem holds exactly in the 
case of a linear equation.
In this paper, we would like to propose 
that this theorem holds to every order
in the perturbative expansion in the case of a linear equation.

In the calculation of the Langevin equation in POM,
we expand the factor $e^{iLQ(t-t_{0})}\frac{1}{1-\Sigma(t,t_{0})}$ 
up to zeroth order in $L_{I}$ and determine the form
of the fluctuation force,
\begin{eqnarray}
\lefteqn{Qe^{iLQ(t-t_{0})}\frac{1}{1-\Sigma(t,t_{0})}iLO(t_{0})} 
&&\nonumber \\
&=& Q{\cal D}(t,t_{0})Q \nonumber \\
&&\times \frac{1}{1+({\cal C}(t,t_{0})-1)Q+({\cal D}(t,t_{0})-1)
+({\cal C}(t,t_{0})-1)Q({\cal D}(t,t_{0})-1)}e^{iL_{0}(t-t_{0})} \nonumber \\
&\sim& Qe^{iL_{0}(t-t_{0})}iLO(t_{0}). \label{eqn:tenkai}
\end{eqnarray}

Now, we confirm that this fluctuation force satisfies the 2nd f-d theorem
employing the following model.
The Hamiltonian is
\begin{eqnarray}
H = \omega_{a}a^{\dagger}a + \int d\omega \omega b^{\dagger}(\omega)b(\omega) 
+ g\int d\omega (a^{\dagger}b(\omega) + ab^{\dagger}(\omega)),
\end{eqnarray}
where both $a$ and $b(\omega)$ are boson operators, which satisfy the 
following commutation relations:
\begin{eqnarray}
[a,a^{\dagger}] = 1,~~~ [b(\omega),b^{\dagger}(\omega')] = \delta(\omega-\omega').
\end{eqnarray}
We consider the case in which $a$ is the system operator 
and $b(\omega)$ the environment operator 
and divide the Hamiltonian into three parts:
\begin{eqnarray}
H_{s} &=& \omega_{a} a^{\dagger}a,           \\
H_{E} &=& \int d \omega \omega b^{\dagger}(\omega)b(\omega),          \\
H_{I} &=& g\int d \omega(a^{\dagger}b(\omega) + ab^{\dagger}(\omega)).
\end{eqnarray}
The projection operator is defined as
\begin{eqnarray}
PO ={}_{b} \langle 0|O|0 \rangle_{b},
\end{eqnarray}
where 
\begin{eqnarray}
b(\omega)|0 \rangle_{b} = 0.
\end{eqnarray}
Substituting these ingredients into Eq. (\ref{eqn:2cd-tcl-flu}), 
we have 
\begin{eqnarray}
\frac{d}{dt}a^{\dagger}(t) = i\omega_{a} a^{\dagger}(t)
-g^2 \int d\omega \frac{e^{i(\omega-\omega_{a})t}-1}{i(\omega-\omega_{a})} a^{\dagger}(t)
+ f(t),
\end{eqnarray}
where
\begin{eqnarray}
f(t) = ig\int d \omega e^{i\omega t}b^{\dagger}(\omega).
\end{eqnarray}
This equation can be rewritten as
\begin{eqnarray}
\frac{d}{dt}a^{\dagger}(t)
&=& i\omega_{a} a^{\dagger}(t)
+ \int^{t}_{0} ds \langle 0|[f(t),f^{\dagger}(s)]|0\rangle  e^{-i\omega_{a}(t-s)}a^{\dagger}(t)
+ f(t) \nonumber \\
&=& i\omega_{a} a^{\dagger}(t)
+ \int^{t}_{0} ds \phi(t-s) e^{-i\omega_{a}(t-s)}a^{\dagger}(t)
+ f(t) \nonumber \\
&=& i\omega_{a} a^{\dagger}(t)
+ \int^{t}_{0} ds \phi(s) e^{-i\omega_{a}s}a^{\dagger}(t)
+ f(t). \label{eqn:shl}
\end{eqnarray}
This equation is the TCL equation, and therefore, we cannot compare this
result with the Mori-Langevin equation, 
which has a time-convolution integral term.
However, if we accept the approximation 
$a^{\dagger}(t-s) \sim e^{-i\omega_{a}s}a^{\dagger}(t)$, 
as in case of the linear harmonic approximation, 
we can conclude that this equation satisfies 
the 2nd f-d theorem to this order of the perturbative expansion.
Therefore, the definition of the fluctuation force (\ref{eqn:tenkai})
is consistent with the 2nd f-d theorem in this case of 
a linear Langevin equation.
We apply this approximate fluctuation force to 
the nonlinear case.

\section{Derivation of a semiclassical equation in IFM} \label{app:fv-qm}

For simplicity, we consider the case of quantum mechanics, in which
the system is denoted by $X$ and the environment by $Q$.
The time evolution of the density matrix is described by 
\begin{eqnarray}
\lefteqn{\rho_{X \cup Q}(x,q;x',q';t) } && \nonumber \\
&&= \int dx_{i}dq_{i}dx'_{i}dq'_{i}U(x,q;x_{i},q_{i};t,t_{i})
\rho_{X \cup Q}(x_{i},q_{i};x'_{i},q'_{i};t_{i})
U^{*}(x',q';x'_{i},q'_{i};t,t_{i}), \nonumber \\
\end{eqnarray}
where
\begin{eqnarray}
U(x,q;x_{i},q_{i};t,t_{i}) &=& \int^{x}_{x_{i}}Dx'
\int^{q}_{q_{i}}Dq' e^{iS(x',q')/\hbar},
\end{eqnarray}
with
\begin{eqnarray}
S(x,q) = \int^{t}_{t_{i}}ds {\cal L}(x(s),q(s)).
\end{eqnarray}
Here, the action $S(x,q)$ is given by
\begin{eqnarray}
S(x,q) &=& S_{X}(x)+S_{Q}(q)+S_{int}(x,q),
\end{eqnarray}
where $S_{X}(x)$ is the action of the system,
$S_{Q}(q)$ the action of the environment, and 
$S_{int}(x,q)$ the action of the interaction part.
As in case of POM, we assume no initial correlation between $X$ and $Q$:
\begin{eqnarray}
\rho_{X \cup Q}(x_{i},q_{i};x'_{i},q'_{i};t_{i}) = \rho_{X}(x_{i},x'_{i};t_{i})
\otimes\rho_{Q}(q_{i},q'_{i};t_{i}).
\end{eqnarray}

The reduced density matrix is defined by tracing out 
the environment degrees of freedom, as 
\begin{eqnarray}
\rho_{r}(x,x';t)
&=& Tr_{(q)}\{ \rho_{X \cup Q}(x,q;x',q';t) \} \nonumber \\
&\sim& \int dx_{i}dx'_{i}\int^{x}_{x_{i}}Dx\int^{x'}_{x'_{i}}Dx'
    e^{i(S_{X}(x) - S_{X}(x'))}e^{iS_{IF}(x,x')}\rho_{X}(x_{i},x'_{i};t) \nonumber \\
&=& \int dx_{i}dx'_{i}\int^{x}_{x_{i}}Dx\int^{x'}_{x'_{i}}Dx'
    e^{iS_{eff}(x,x')}\rho_{X}(x_{i},x'_{i};t),
\end{eqnarray}
where the functions $e^{iS_{IF}(x,x')}$ and $S_{eff}(x,x')$ 
are the influence functional and the effective action, respectively.
It is difficult to trace out the environment degrees of freedom
completely, and therefore, we have carried out a perturbative expansion 
in going from the first line to the second line.
As an example, we consider the interaction
\begin{eqnarray}
{\cal L}
_{int}(x,q) = x\Xi(q).
\end{eqnarray}
The influence action $S_{IF}(x,x')$ is expanded up to second order in 
powers of $x$:
\begin{eqnarray}
\lefteqn{S_{IF}[x,x'] }&&\nonumber \\
&=& \int^{t}_{t_{i}}ds F(s)[x(s) - x'(s)] \nonumber \\
&&\hspace*{-0.5cm}  +\frac{1}{2}\int^{t}_{t_{i}}ds ds'
  \Bigg[\Big(R(s,s')+iI(s,s')\Big)x(s)x(s') + \Big(-R(s,s')+iI(s,s')\Big)x'(s)x'(s') \nonumber \\ &&\hspace*{-0.5cm}   + \Big(R(s,s')-iI(s,s')\Big)x(s)x'(s') + \Big(-R(s,s')-iI(s,s')\Big)x'(s)x(s')\Bigg], 
\end{eqnarray}
where $F(s)$, $R(s,s')$ and $I(s,s')$ are real (Hermitian) functions.
The effective action $S_{eff}$ is a complex function.
To make it real, we introduce a new variable $\xi$,
and with it we obtain the stochastic effective action,
\begin{eqnarray}
\tilde{S}_{eff}(x,x',\xi) 
&=& S_{X}(x)-S_{X}(x') + {\rm Re}(S_{IF}(x,x'))+\int^{t}_{t_{i}}ds \xi(s)[x(s)-x'(s)] \nonumber \\
&=& S_{X}(x)-S_{X}(x') + \tilde{S}_{IF}(x,x',\xi).
\end{eqnarray}
This is derived by using the relation
\begin{eqnarray}
e^{i/\hbar S_{eff}(x,x')}
\equiv \int D\xi P[\xi]e^{i/\hbar \tilde{S}_{eff}(x,x',\xi)}, \label{eqn:gauss-rel}
\end{eqnarray}
where $P[\xi]$ is the stochastic weight defined by
\begin{eqnarray}
P[\xi] = \frac{1}{N}\exp
\left(
-\frac{1}{2\hbar}\int^{t}_{t_{i}}ds_{1}ds_{2}
\xi(s_{1})I^{-1}(s_{1},s_{2})\xi(s_{2})
\right). \label{eqn:swqm}
\end{eqnarray}
In deriving this relation, we have assumed that it is possible to 
carry out the Gaussian integral.
(See Appendix C for more details.)

If the system behaves quasi-classically,
the reduced density matrix becomes nearly diagonal.
Then, the major contributing path is obtained
by applying a variational principle to the stochastic effective action:
\begin{eqnarray}
\left.
\frac{\delta \tilde{S}_{eff}(\bar{x},\Delta,\xi)}{\delta \Delta} \right|_{\Delta=0}
&=& 0,
\end{eqnarray}
where
\begin{eqnarray}
\bar{x} &=& \frac{x+x'}{2}, \\
\Delta &=& x-x'.
\end{eqnarray}
The equation derived in this manner is a semiclassical equation 
that includes quantum effects of the environment:
\begin{eqnarray}
\frac{\delta S_{X}(\bar{x})}{\delta \bar{x}(s)} + F(s)
+\int^{s}_{t_{i}}ds' R(s,s')\bar{x}(s') = -\xi(s).  \label{eqn:sto-e.o.m.}
\end{eqnarray}
Here, we assume that the new variable $\xi$ can be interpreted as
a fluctuation force that is distributed with the stochastic weight $P[\xi]$
defined in Eq.~(\ref{eqn:swqm}).
Then, Eq.~(\ref{eqn:sto-e.o.m.}) can be regarded as a kind of 
Langevin equation.

\section{Gaussian integral}

To confirm Eq. (\ref{eqn:gauss-rel}), we calculate the following integral: 
\begin{eqnarray}
\lefteqn{\int D\xi P[\xi]e^{(i/\hbar) \tilde{S}_{IF}(x,x',\xi)}} && \nonumber \\
&=& \int D\xi \frac{1}{N}\exp
\Big\{
\frac{i}{\hbar}{\rm Re} S_{IF}(x,x')
+ \frac{i}{\hbar}\int^{t}_{t_{i}} ds \xi(s)(x(s)-x'(s)) \nonumber \\
&& -\frac{1}{2\hbar}\int^{t}_{t_{i}} ds_{1} ds_{2}
\xi(s_{1})I^{-1}(s_{1},s_{2})\xi(s_{2})
\Big\} \nonumber \\
&=& \int D\xi \frac{1}{N}\exp \left\{
\frac{i}{\hbar}{\rm Re} S_{IF}(x,x')
+ \frac{i}{\hbar} \xi_{i}(x_{i}-x'_{i})
-\frac{1}{2\hbar}\xi_{i}I^{-1}_{ij}\xi_{j}
\right\}. \label{eqn:gauss-dis}
\end{eqnarray}
Here, we omit the summation symbol.
We assume that the matrix $I^{-1}$ can be 
diagonalized by a unitary matrix $U$:
\begin{eqnarray}
UI^{-1}U^{\dagger}
=
\left(
\begin{array}{lll}
\lambda_{1} &  & \\
       & \lambda_{2} & \\
       &             &  \ddots
\end{array}
\right).
\end{eqnarray}
The terms that appear in the brackets on the r.h.s. of 
Eq.~(\ref{eqn:gauss-dis}) can be transformed as
\begin{eqnarray}
\frac{i}{\hbar} \xi_{i}(x_{i}-x'_{i})
-\frac{1}{2\hbar}\xi_{i}I^{-1}_{i,j}\xi_{j}
&=&
-\frac{\lambda_{i}}{2\hbar}\eta_{i}^2
+ \frac{i}{\hbar}\eta_{i}(y_{i}-y'_{i})           \nonumber \\
&=& -\frac{\lambda_{i}}{2\hbar}\{ \eta_{i}^2 -\frac{2i}{\lambda_{i}}\eta_{i}(y_{i}-y'_{i}) \} \nonumber \\
&=& -\frac{\lambda_{i}}{2\hbar}
\{ (\eta_{i}-\frac{i}{\lambda_{i}}(y_{i}-y'_{i}))^2
+\frac{1}{\lambda^2_{i}}(y_{i}-y'_{i})^2 \},\nonumber \\
\end{eqnarray}
where $(Ux)_{i}=y_{i},(U\xi)_{i}=\eta_{i}$.
Furthermore, we assume that all the eigenvalues $\lambda_{i}$ are positive.
Then, we can carry out the Gaussian integral:
\begin{eqnarray}
\int D\xi \exp
\{ -\frac{\lambda_{i}}{2\hbar}(\eta_{i}-\frac{i}{\lambda_{i}}(y_{i}-y'_{i}))^2 \}
&=& \prod_{i} \int^{\infty}_{-\infty} d\eta_{i} \exp
\left\{ -\frac{\lambda_{i}}{2\hbar}\eta_{i}^2 \right\} \nonumber \\
&=& \prod_{i}\sqrt{\frac{2\hbar \pi}{\lambda_{i}}} \nonumber \\
&=& [det (2\hbar \pi I_{ij})]^{1/2}.
\end{eqnarray}
It follows that 
\begin{eqnarray}
\int D\xi P[\xi]e^{(i/\hbar) \tilde{S}_{IF}(x,x',\xi)}
&=& \exp \left\{
\frac{i}{\hbar}{\rm Re} S_{IF}(x,x')
-\frac{1}{2\hbar\lambda_{i}}(y_{i}-y'_{i})^2
\right\} \nonumber \\
&=& \exp \frac{i}{\hbar} \left\{
{\rm Re} S_{IF}(x,x')
+\frac{i}{2}(x_{i}-x'_{i})I_{ij}(x_{j}-x'_{j})
\right\} \nonumber \\
&=& e^{(i/\hbar) S_{IF}(x,x')},
\end{eqnarray}
where
\begin{eqnarray}
N = [Det (2\hbar\pi I)]^{1/2}.
\end{eqnarray}
Thus, the relation (\ref{eqn:gauss-rel}) is confirmed. 
Equation (\ref{eqn:phi-gaussf}) is 
the quantum field theoretical version of this relation.


\newpage

\vspace{2cm}

\end{document}